\documentclass[sigconf, 10pt]{acmart}

\usepackage{booktabs}
\usepackage{xcolor}
\usepackage{graphicx}
\usepackage{multirow}
\usepackage{subcaption}
\usepackage{amsmath}
\usepackage{verbatim}
\usepackage[ruled,linesnumbered]{algorithm2e}
\usepackage{threeparttable}

\copyrightyear{2019} 
\acmYear{2019} 
\setcopyright{acmcopyright}
\acmConference[SIGMOD '19]{2019 International Conference on Management of Data}{June 30-July 5, 2019}{Amsterdam, Netherlands}
\acmBooktitle{2019 International Conference on Management of Data (SIGMOD '19), June 30-July 5, 2019, Amsterdam, Netherlands}
\acmPrice{15.00}
\acmDOI{10.1145/3299869.3319888}
\acmISBN{978-1-4503-5643-5/19/06}

\acmArticle{4}
\acmPrice{15.00}

\newcommand{\yell}[1]{{\color{red}#1}}

\begin{document}

\title{HoloDetect: Few-Shot Learning for Error Detection}

\author{Alireza Heidari$^*$, Joshua McGrath$^\dagger$, Ihab F. Ilyas$^*$, Theodoros Rekatsinas$^\dagger$}
\email{a5heidar@uwaterloo.ca, mcgrath@cs.wisc.edu, ilyas@uwaterloo.ca, thodrek@cs.wisc.edu}
\affiliation{%
  \institution{$^*$University of Waterloo and $^\dagger$University of Wisconsin - Madison}
}

\renewcommand{\shortauthors}{Heidari, McGrath, Ilyas, and Rekatsinas}
\renewcommand{\authors}{Heidari, McGrath, Ilyas, and Rekatsinas}

\begin{abstract}
We introduce a few-shot learning framework for error detection. We show that data augmentation (a form of weak supervision) is key to training high-quality, ML-based error detection models that require minimal human involvement. Our framework consists of two parts: (1) an expressive model to learn rich representations that capture the inherent syntactic and semantic heterogeneity of errors; and  (2) a data augmentation model that, given a small seed of clean records, uses dataset-specific transformations to automatically generate additional training data. Our key insight is to learn data augmentation policies from the noisy input dataset in a weakly supervised manner. We show that our framework detects errors with an average precision of \textasciitilde94\% and an average recall of \textasciitilde93\% across a diverse array of datasets that exhibit different types and amounts of errors. We compare our approach to a comprehensive collection of error detection methods, ranging from traditional rule-based methods to ensemble-based and active learning approaches. We show that data augmentation yields an average improvement of 20 $F_1$ points while it requires access to 3$\times$ fewer labeled examples compared to other ML approaches.
\end{abstract}


\maketitle

\section{Introduction}
\label{sec:introduction}

Error detection is a natural first step in every data analysis pipeline~\cite{IlyasC15,osborne2013best}. Data inconsistencies due to incorrect or missing data values can have a severe negative impact on the quality of downstream analytical results. However, identifying errors in a noisy dataset can be a challenging problem. Errors are often heterogeneous and exist due to a diverse set of reasons (e.g., typos, integration of stale data values, or misalignment), and in many cases can be rare. This makes manual error detection prohibitively time consuming.

Several error detection methods have been proposed in the literature to automate error detection~\cite{Rahm00,IlyasC15,Fan2012,HalevyBook}. Most of the prior works leverage the side effects of data errors to solve error detection. For instance, many of the proposed methods rely on violations of integrity constraints~\cite{IlyasC15} or value-patterns~\cite{wrangler} or duplicate detection~\cite{Elmagarmid07,2010Naumann} and outlier detection~\cite{Dasu2012,Wu:2013,2015combining} methods to identify erroneous records. While effective in many cases, these methods are tailored to specific types of side effects of erroneous data. As a result, their recall for identifying errors is limited to errors corresponding to specific side effects (e.g., constraint violations, duplicates, or attribute/tuple distributional shifts)~\cite{AbedjanCDFIOPST16}.

One approach to address the heterogeneity of errors and their side effects is to combine different detection methods in an ensemble~\cite{AbedjanCDFIOPST16}. For example, given access to different error detection methods, one can apply them sequentially or can use voting-based ensembles to combine the outputs of different methods. Despite the simplicity of ensemble methods, their performance can be sensitive to how different error detectors are combined~\cite{AbedjanCDFIOPST16}. This can be either with respect to the order in which different methods are used or the confidence-level associated with each method. Unfortunately, appropriate tools for tuning such ensembles are limited, and the burden of tuning these tools is on the end-user.

A different way to address heterogeneity is to cast error detection as a machine learning (ML) problem, i.e., a binary classification problem: given a dataset, classify its entries as erroneous or correct. One can then train an ML model to discriminate between erroneous and correct data. Beyond automation, a suitably expressive ML model should be able to capture the inherent heterogeneity of errors and their side effects and will not be limited to low recall. However, the end-user is now burdened with the collection of enough labeled examples to train such an expressive ML model. 

\subsection{Approach and Technical Challenges}
\label{sec:approach}

 We propose a few-shot learning framework for error detection based on weak supervision~\cite{Ratner:2017:SRT:3173074.3173077,DBLP:conf/kdd/Re18}, which exploits noisier or higher-level signals to supervise ML systems. We start from this premise and show that {\em data augmentation}~\cite{Perez2017TheEO,45820}, a form of weak supervision, enables us to train high-quality ML-based error detection models with minimal human involvement. 

Our approach exhibits significant improvements over a comprehensive collection of error detection methods: we show that our approach is able to detect errors with an average precision of \textasciitilde94\% and an average recall of \textasciitilde93\%, obtaining an average improvement of 20 $F_1$ points against competing error detection methods. At the same time, our weakly supervised methods require access to 3$\times$ fewer labeled examples compared to other ML approaches. Our ML-approach also needs to address multiple technical challenges:
\begin{itemize}

\item {\bf [Model]}\ \  The heterogeneity of errors and their side effects makes it challenging to identify the appropriate statistical and integrity properties of the data that should be captured by a model in order to discriminate between erroneous and correct cells. These properties correspond to attribute-level, tuple-level, and dataset-level features that describe the distribution governing a dataset. Hence, we need an appropriately expressive model for error detection that captures all these properties (features) to maximize recall. 

\item {\bf [Imbalance]} \ \ Often, errors in a dataset are limited. ML algorithms tend to produce unsatisfactory classifiers when faced with imbalanced datasets. The features of the minority class are treated as noise and are often ignored. Thus, there is a high probability of misclassification of the minority class as compared to the majority class. To deal with imbalance, one needs to develop strategies to balance classes in the training data. Standard methods to deal with the imbalance problem such as resampling  can be ineffective due to error heterogeneity as we empirically show in our experimental evaluation. 

\item {\bf [Heterogeneity] } Heterogeneity  amplifies the imbalance problem as certain errors and their side effects can be underrepresented in the training data. Resampling the training data does not ensure that errors with different properties are revealed to the ML model during training. While active learning can help counteract this problem in cases of moderate imbalance~\cite{chawla2004special,Ertekin:2007:ALC:1277741.1277927}, it tends to fail in the  case of extreme imbalance~\cite{He:2013:ILF:2559492}  (as in the case of error detection). This is because the lack of labels prevents the selection scheme of active learning from identifying informative instances for labeling~\cite{He:2013:ILF:2559492}. Different methods that are robust to extreme imbalance are needed.
\end {itemize}

A solution that addresses the aforementioned challenges needs to: (1)~introduce an expressive model for error detection, while avoiding explicit feature engineering; and (2)~propose novel ways to handle the extreme imbalance and heterogeneity of data in a unified manner.

\subsection{Contributions and Organization}
\label{sec:contrib}

 To obviate the need for feature engineering we design a representation learning framework for error detection. To address the heterogeneity and imbalance challenges we introduce a data augmentation methodology for error detection. We summarize the main contributions as follows:
\begin{itemize}
\item We introduce a template ML-model to learn a representation that captures attribute-, tuple-, and dataset-level features that describe a dataset. We demonstrate that representation learning obviates the need for feature engineering. Finally, we show via ablation studies that all granularities need to be captured by error detection models to obtain high-quality results.

\item We show how to use data augmentation to address data imbalance. Data augmentation proceeds as follows: Given a small set of labeled data, it allows us to generate synthetic examples or errors by transforming correct examples in the available training data. This approach minimizes the amount of manually labeled examples required. We show that in most cases a small number of labeled examples are enough to train high-quality error detection models.

\item We present a weakly supervised method to learn data transformations and data augmentation policies (i.e., the distribution over those data transformation) directly from the noisy input  dataset. The use of different transformations during augmentation provides us with examples that correspond to different types of errors, which enables us to address the aforementioned heterogeneity challenge. 

\end{itemize}
The remainder of the paper proceeds as follows: In Section~\ref{sec:background} we review background concepts. Section~\ref{sec:overview} provides an overview of our weak supervision framework. In Section~\ref{sec:model}, we introduce our representation learning approach to error detection. In Section~\ref{sec:weaksup}, we establish a data augmentation methodology for error detection, and in Section~\ref{sec:experiments}, we evaluate our proposed solutions. We discuss related work in Section~\ref{sec:related} and summarize key points of the paper in Section~\ref{sec:conclusions}.
\section{Background}
\label{sec:background}
We review basic background material for the problems and techniques discussed in this paper.

\subsection{Error Detection}
\label{sec:bgerror}
The goal of error detection is to identify incorrect entries in a dataset. Existing error detection methods can be categorized in three main groups: (1) Rule-based methods~\cite{holistic, dallachiesa2013nadeef} rely on integrity constraints such as functional dependencies and denial constraints, and suggest errors based on the violations of these rules. Denial Constraints (DCs) are first order logic formulas that subsume several types of integrity constraints~\cite{Chomicki:2005:MIM:1709465.1709573}. Given a set of operators $B = \{=, <, >, \neq, \leq, \approx \}$, with $\approx$ denoting similarity, DCs take the form $\forall t_i, t_j \in D:  \neg (P_1 \wedge \dots \wedge P_k \wedge \dots \wedge P_K)$ where $D$ is a dataset with attributes $A = \{A_1, A_2, \dots, A_N\}$, $t_i$ and $t_j$ are tuples, and each predicate $P_k$ is of the form $(t_i[A_n] \operatorfont{~op~} t_j[A_{m}])$ or $(t_i[A_n] \operatorfont{~op~} \alpha)$ where $A_n, A_m \in A$, $\alpha$ is a constant and $\operatorfont{~op~} \in B$. 
 (2) Pattern-driven methods leverage normative syntactic patterns and identify erroneous entries such as those that do not conform with these patterns~\cite{wrangler}. (3) Quantitative error detection focuses on outliers in the data and declares those to be errors~\cite{hellerstein2008quantitative}. A problem related to error detection is record linkage~\cite{Elmagarmid07,2010Naumann,HalevyBook}, which tackles the problem of identifying if multiple records refer to the same real-world entity. While it can also be viewed as a classification problem, it does not detect errors in the data and is not the focus of this paper.

\subsection{Data Augmentation}
\label{sec:bgAugment}
Data augmentation is a form of weak supervision~\cite{DBLP:conf/kdd/Re18} and refers to a family of techniques that aim to extend a dataset with additional data points. Data augmentation is typically applied to training data as a way to reduce overfitting of models~\cite{45820}. Data augmentation methods typically consist of two components: (1) a set of {\em data transformations} that take a data point as input and generate an altered version of it, and (2) an {\em augmentation policy} that determines how different transformations should be applied, i.e., a distribution over different transformations. Transformations are typically specified by domain experts while policies can be either pre-specified~\cite{Perez2017TheEO} or learned via reinforcement learning or random search methods~\cite{cubuk2018autoaugment,DBLP:conf/nips/RatnerEHDR17}. In contrast to prior work, we show that for error detection both transformations and policies can be learned directly from the data.

\subsection{Representation Learning} 
\label{sec:bgRepLearn}
The goal of representation learning is to find an appropriate representation of data (i.e., a set of features) to perform a machine learning task~\cite{Bengio:2013:RLR:2498740.2498889}. In our error detection model we build upon three standard representation learning techniques:

\vspace{3pt}\noindent\textbf{Neural Networks} Representation learning is closely related to neural networks~\cite{GoodBengCour16}. The most basic neural network takes as input a vector $\mathbf{x}$ and performs an affine transformation of the input $\mathbf{w}\mathbf{x} + b$. It also applies a non-linear activation function $\sigma$ (e.g., a sigmoid) to produce the output $\sigma(\mathbf{w}\mathbf{x} + b)$. Multiple layers can be stacked together to create more complex networks. In a neural network, each hidden layer maps its input data to an internal representation that tends to capture a higher level of abstraction.

\vspace{3pt}\noindent\textbf{Highway Neural Networks} Highway Networks, adapt the idea of having ``shortcut'' gates that allow unimpeded information to flow across non-consecutive layers~\cite{srivastava2015highway}. Highway Networks are used to improve performance in many domains such as speech recognition~\cite{zhang2016highway} and language modeling~\cite{kim2016character}, and their variants called Residual networks have been useful for many computer vision problems~\cite{he2016deep}

\vspace{3pt}\noindent\textbf{Distributed Representations} Distributed representations of symbolic data~\cite{Hinton:1986:DR:104279.104287} were first used in the context of statistical language model~\cite{Bengio:2003:NPL:944919.944966}. The goal here is to learn a mapping of a token (e.g., a word) to a vector of real numbers, called a {\em word embedding}. Methods to generate these mappings include neural networks~\cite{word2vec}, dimensionality reduction techniques such as PCA~\cite{lebret-collobert:2014:EACL}, and other probabilistic techniques~\cite{Globerson:2007:EEC:1314498.1314572}.

\section{Framework Overview}
\label{sec:overview}
\begin{figure*}
  \centering
  \includegraphics[width=\textwidth]{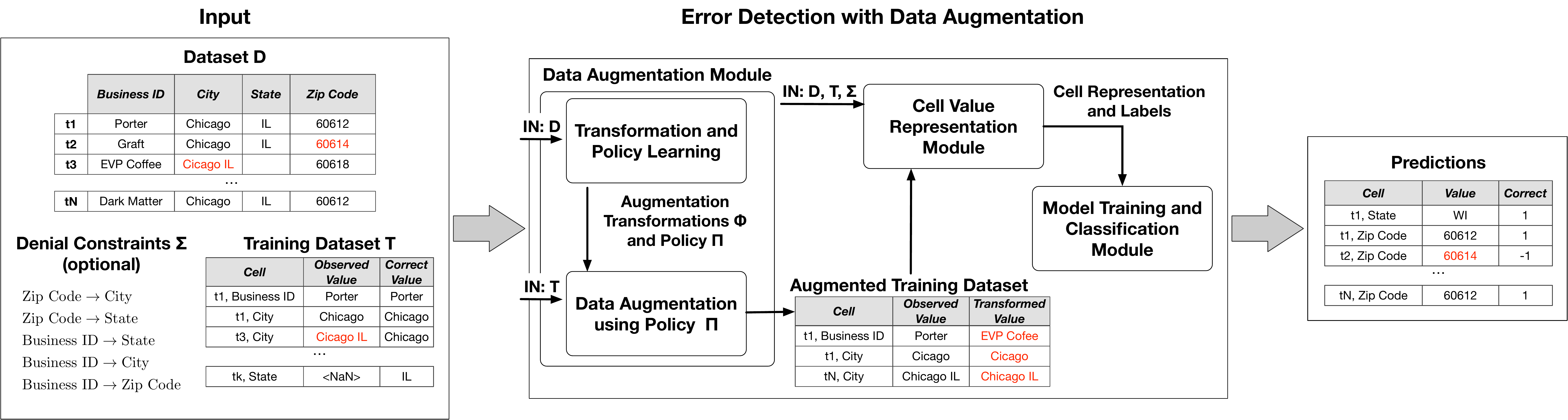}
  \caption{Overview of Error Detection with Augmentation.}
  \label{fig:overview}
\end{figure*}
We formalize the problem of error detection and provide an overview of our solution to error detection.

\subsection{Problem Statement}
\label{sec:prob}
The goal of our framework is to identify erroneous entries in a relational dataset $D$. We de{}note $A = \{A_1, A_2, \dots, A_N\}$ the attributes of $D$. We follow set semantics and consider $D$ to be a set of tuples. Each tuple $t \in D$ is a collection of cells $C_t = \{t[A_1], t[A_2], \dots, t[A_N]\}$ where $t[A_i]$ denotes the value of attribute $A_i$ for tuple $t$. We use $C_D$ to denote the set of cells contained in $D$. The input dataset $D$ can also be accompanied by a set of integrity constraints $\Sigma$, such as Denial Constraints as described in Section~\ref{sec:bgerror}.

 We assume that errors in $D$ appear due to inaccurate cell assignments. More formally, for a cell $c$ in $C_D$ we denote by $v^*_c$ its unknown true value and $v_c$ its observed value. We define an {\em error} in D to be each cell $c$ with $v_c \neq v^*_c$. We define a {\em training dataset} $T$ to be a set of tuples $T = \{(c, v_c, v^*_c)\}_{c \in C_T}$ where $C_T \subset C_D$. $T$ provides labels (i.e., correct or erroneous) for a subset of cells in $D$. We also define a variable $E_c$ for each cell $c \in C_D$ with $E_c = -1$ indicating that the cell is erroneous and with $E_c = 1$ indicating that the cell is correct. For each $E_c$ we denote $e^*_c$ its unknown true assignment. 

 Our goal is stated as follows: given a dataset $D$ and a training dataset $T$ find the most probable assignment $\hat{e}_c$ to each variable $E_c$ with $c \in C_D \setminus C_T$. We say that a cell is correctly classified as erroneous or correct when $\hat{e}_c = e^*_c$.

\subsection{Model Overview} 
\label{sec:models}

 
Prior models for error detection focus on specific side effects of data errors. For example, they aim to detect errors by using only the violations of integrity constraints or aim to identify outliers with respect to the data distribution that are introduced due to errors. Error detectors that focus on specific side effects, such as the aforementioned ones, are not enough to detect errors with a high recall in heterogeneous datasets~\cite{Abedjan_2016}. This is because many errors may not lead to violations of integrity constraints, nor appear as outliers in the data. We propose a different approach: we model the process by which the entries in a dataset are generated, i.e., we model the distribution of both correct and erroneous data. This approach enables us to discriminate better between these two types of data.


We build upon our recent Probabilistic Unclean Databases (PUDs) framework that introduces a probabilistic framework for managing noisy relational data~\cite{puds}. We follow the abstract generative model for noisy data from that work, and introduce an instantiation of that model to represent the distribution of correct and erroneous cells in a dataset.

We consider a noisy channel model for databases that proceeds in two steps: First, a clean database is sampled from a probability distribution $I^*$. Distribution $I^*$ captures how values within an attribute and across attributes are distributed and also captures the compatibility of different tuples (i.e., it ensures that integrity constraints are satisfied). To this end, distribution $I^*$ is defined over attribute-, tuple-, and dataset-level features of a dataset. Second, given a clean database sampled by $I^*$, errors are introduced via a noisy channel that is described by a conditional probability distribution $R^*$. Given this model, $I^*$ characterizes the probability of the unknown true value $P(v^*_c)$  of a cell $c$ and $R^*$ characterizes the conditional probability $P(v_c|v^*_c)$ of its observed value. Distribution $I^*$ is such that errors in dataset $D$ lead to low probability instances. For example, $I^*$ assigns zero probability to datasets with entries that lead to constraint violations. 

The goal is to learn a representation that captures the distribution of the correct cells ($I^*$) and how errors are introduced ($R^*$). Our approach relies on learning two models:

\vspace{3pt}\noindent(1)  {\it Representation Model} We learn a representation model $Q$ that approximates distribution $I^*$ on the attribute, record, and dataset level. We require that $Q$ is such that the likelihood of correct cells given $Q$ will be high, while the likelihood of erroneous cells given $Q$ is low. This property is necessary for a classifier $M$ to discriminate between correct and erroneous cells when using representation $Q$. We rely on representation learning techniques to learn $Q$ jointly with $M$.  

\vspace{3pt}\noindent(2) {\it Noisy Channel} We learn a generative model $H$ that approximates distribution $R^*$. This model consists of a set of transformations $\Phi$ and a policy $\Pi$. Each transformation $\phi \in \Phi$ corresponds to a function that takes as input a cell $c$ and transforms its original value $v_c$ to a new value $v^{\prime}_c$, i.e., $\phi(v_c) = v^{\prime}_c$. Policy $\Pi$ is defined as a conditional distribution $P(\Phi|v_c)$. As we describe next, we use this model to generate training data---via data augmentation---for  learning $Q$ and $M$.

%

We now present the architecture of our framework. The  modules described next are used to learn the noisy channel $H$, perform data augmentation by using $H$, and learn the representation model $Q$ jointly with a classifier $M$ that is used to detect errors in the input dataset.

\begin{figure*}
  \centering
  \includegraphics[width=\textwidth]{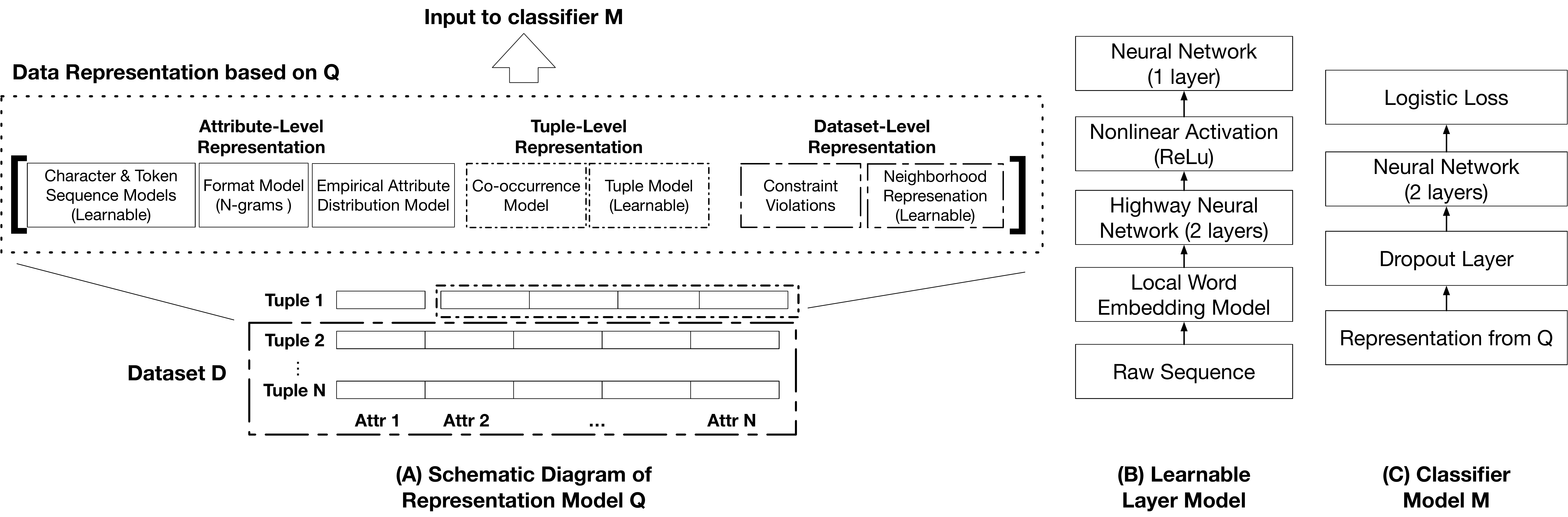}
  \caption{(A) A diagram of the representation model Q. Models associated learnable layers that are jointly trained with classifier $M$. (B) Architecture diagram of the learnable layers in Q. (C) The architecture of classifier M.}
  \label{fig:models_overview}
\end{figure*}

\subsection{Framework Overview}
\label{sec:sol_overview}
Our framework takes as input a noisy dataset $D$, a training dataset $T$, and (optionally) a set of denial constraints $\Sigma$. To learn $H$, $Q$, and $M$ from this input we use three core modules:

\vspace{3pt}\noindent\textbf{Module 1: Data Augmentation} This module learns the noisy channel $H$ and uses it to generate additional training examples by transforming some of the labeled examples in $T$. The output of this module is a set of additional examples $T_H$. The operations performed by this module are:

\vspace{3pt}\noindent(1) {\it Transformation and Policy Learning:} The goal here is to learn the set of transformations $\Phi$ and the policy $\Pi$ that follow the data distribution in $D$. We introduce a weakly supervised algorithm to learn $\Phi$ and $\Pi$. This algorithm is presented in Section~\ref{sec:weaksup}.
	
\vspace{3pt}\noindent(2) {\it Example Generation:} Given transformations $\Phi$ and policy $\Pi$, we generate a set of new training examples $T_H$ that is combined with $T$ to train the error detection model. To ensure high-quality training data, this part augments only cells that are marked correct in $T$. Using this approach, we obtain a balanced training set where examples of errors follow the distribution of errors in $D$. This is because transformations are chosen with respect to policy $\Pi$ which is learned from $D$.

\vspace{3pt}\noindent\textbf{Module 2: Representation} This module combines different representation models to form model $Q$. Representation $Q$ maps a cell values $v_c$ to to a fixed-dimension real-valued vector $f_c \in R^d$. To obtain $f_c$ we concatenate the output of different representation models, each of which targets a specific context (i.e., attribute, tuple, or dataset context).

We allow a representation model to be learned during training, and thus, the output of a representation model can correspond to a vector of variables (see Section~\ref{sec:model}). For example, the output of a representation model can be an embedding $u_c$ obtained by a neural network that is learned during training or may be fixed to the number of constraint violations value $v_c$ participates in.

\vspace{3pt}\noindent\textbf{Module 3: Model Training and Classification} This module is responsible for training a classifier $M$ that given the representation of a cell value determines if it is correct or erroneous, i.e., $M: R^d \rightarrow \{\text{``correct'' (+1)}, \text{``error (-1)''}\}$. During training, the classifier is learned by using both the initial training data $T$ and the augmentation data $T_A$. At prediction time, the classifier $M$ takes as input the cell value representation for all cells in $D \setminus T$ and assigns them a label from $\{\text{``correct''}, \text{``error''}\}$ (see Section~\ref{sec:model}).

An overview of how the different modules are connected is shown in Figure~\ref{fig:overview}. First, Module 1 learns transformations $\Phi$ and policy $\Pi$. Then, Module 2 grounds the representation model $Q$ of our error detection model. Subsequently, $Q$ is connected with the classifier model $M$ in Module 3 and trained jointly. The combined model is used for error detection.
\section{Representations of Dirty Data}
\label{sec:model}

We describe how to construct the representation model $Q$ (see Section~\ref{sec:models}). We also introduce the classifier model $M$, and describe how we train $Q$ and $M$. 

\subsection{Representation Models}
To approximate the data generating distribution $I^*$, the model $Q$ needs to capture statistical characteristics of cells with respect to attribute-level, tuple-level, and dataset-level contexts. An overview of model $Q$ is shown in Figure~\ref{fig:models_overview}(A). As shown, $Q$ is formed by concatenating the outputs of different models.  Next, we review the representation models we use for each of the three contexts. The models introduced next correspond to a bare-bone set that captures all aforementioned contexts, and is currently implemented in our prototype. More details on our implementation are provided in Appendix~\ref{sec:model_details}. Our architecture can trivially accommodate additional models or more complex variants of the current models.

\vspace{3pt}\noindent{\bf Attribute-level Representation:} Models for this context capture the distributions governing the values and format for an attribute. Separate models are used for each attribute $A_i$ in dataset $D$. We consider three types of models: (1) {\em Character and token sequence models} that capture the probability distribution over sequences of characters and tokens in cell values. These models correspond to learnable representation layers. Figure~\ref{fig:models_overview}(B) shows the deep learning architecture we used for learnable layers. (2) {\em Format models} that capture the probability distribution governing the format of the attribute. In our implementation, we consider an n-gram model that captures the format sequence over the cell value. Each n-gram is associated with a probability that is learned directly from dataset $D$. The probabilities are aggregated to a fixed-dimension representation by taking the probabilities associated with the least-$k$ probable n-grams. (3) {\em Empirical distribution models} that capture the empirical distribution of the attribute associated with a cell. These can be learned directly from the input dataset $D$. The representation here is a scalar that is the empirical probability of the cell value.

\vspace{3pt}\noindent{\bf Tuple-level Representation:} Models for this context capture the joint distribution of different attributes. We consider two types of models: (1) {\em Co-occurrence models} that capture the empirical joint distribution over pairs of attributes. (2) A learnable {\em tuple representation}, which captures the joint distribution across attributes given the observed cell value. Here, we first obtain an embedding of the tuple by following standard techniques based on word-embedding models~\cite{Q17-1010}. These embeddings are passed through a learnable representation layer (i.e., a deep network) that corresponds to an additional non-linear transform (see Figure~\ref{fig:models_overview}(B)). For co-occurrence, we learn a single representation for all attributes. For tuple embeddings, we learn a separate model per attribute.

\vspace{3pt}\noindent{\bf Dataset-level Representation:} Models for this context capture a distribution that governs the compatibility of tuples and values in the dataset $D$. We consider two types of models: (1) {\em Constraint-based models} that leverage the integrity constraints in $\Sigma$ (if given) to construct a representation model for this context. Specifically, for each constraint $\sigma \in \Sigma$ we compute the number of violations associated with the tuple of the input cell. (2) A {\em neighborhood-based representation} of each cell value that is informed by a dataset-level embedding of $D$ transformed via a learnable layer. Here, we train a standard word-embedding model where each tuple in $D$ is considered to be a document. To ensure that the embeddings are not affected by the sequence of values across attributes we extend the context considered by word-embeddings to be the entire tuple and treat the tuple as a bag-of-words. These embeddings are given as input to a learnable representation layer that follows the architecture in Figure~\ref{fig:models_overview}(B).

The outputs of all models are concatenated into a single vector that is given as input to Classifier $M$. Learnable layers are trained jointly with $M$. To achieve high-quality error detection, features from all contexts need to be combined to form model $Q$. In Section~\ref{sec:experiments}, we present an ablation study which demonstrates that all features from all types of contexts are necessary to achieve high-quality results.

\subsection{Error Classification}
\label{sec:classification}
The classifier $M$ of our framework corresponds to a two-layer fully-connected neural network, with a ReLU activation layer, and  followed by a {\it Softmax} layer. The architecture of $M$ is shown in Figure~\ref{fig:models_overview}(C). Given the modular design of our architecture, Classifier $M$ can be easily replaced with other models. Classifier $M$ is jointly trained with the representation model $Q$ by using the training data in $T$ and the data augmentation output $T_H$. We use ADAM~\cite{journals/corr/KingmaB14} to train our end-to-end model.

More importantly, we calibrate the confidence of the predictions of $M$ using {\it Platt Scaling}~\cite{DBLP:conf/icml/GuoPSW17,PlattProbabilisticOutputs1999} on a holdout-set from the training data $T$ (i.e., we keep a subset of $T$ for calibration). Platt Scaling proceeds as follows: Let $z_i$ be the score for class $i$ output by $M$. This score corresponds to non-probabilistic prediction. To convert it to a calibrated probability, Platt Scaling learns scalar parameters $a, b \in R$ and outputs $\hat{q}_i = \sigma(a z_i + b)$ as the calibrated probability for prediction $z_i$. Here, $\sigma$ denotes the sigmoid function. Parameters $a$ and $b$ are learned by optimizing the negative log-likelihood loss over the holdout-set. It is important to note that the parameters of $M$ and $Q$ are fixed at this stage.

\section{Data Augmentation Learning}
\label{sec:weaksup}

Having established a representation model $Q$ for the data generating distribution $I^*$, we now move to modeling the noisy channel distribution $R^*$. We assume the noisy channel can be specified by a set of transformation functions $\Phi$ and a policy $\Pi$ (i.e., a conditional distribution over $\Phi$ given a cell value). Our goal is to learn $\Phi$ and $\Pi$ from few example errors and use it to generate training examples to learn model $Q$. 

\subsection{Noisy Channel Model}
\label{sec:nc_model}

We aim to limit the number of manually labeled data required for error detection. Hence, we consider a simple noisy channel model that can be learned from few and potentially noisy training data. Our noisy channel model treats cell values as strings and introduces errors to a clean cell value $v^*$ by applying a transformation $\phi$ to obtain a new value $v = \phi(v^*)$. We consider that each function $\phi \in \Phi$ belongs to one of the following three templates:
\begin{itemize}
\item Add characters: $ \varnothing\longmapsto [a-z]^+$
\item Remove characters: $[a-z]^+ \longmapsto \varnothing$
\item Exchange characters: $[a-z]^+ \longmapsto [a-z]^+$ (the left side and right side are different)
\end{itemize}

Given these templates, we assume that the noisy channel model introduces errors via the following generative process: Given a clean input value $v^*$, the channel samples a transformation $\phi$ from a conditional distribution $\Pi(v^*) = P(\Phi|v^*)$, i.e., $\phi \sim \Pi(v^*)$ and applies $\phi$ once to a substring or position of the input cell value. We refer to $\Pi$ as a {\em policy}. If the transformation $\phi$ can be applied to multiple positions or multiple substrings of $v^*$ one of those positions or strings is selected uniformly at random. 

For example, to transform Zip Code ``60612'' to ``606152'', the noisy channel model we consider can apply the exchange character function $T:60612\longmapsto 606152$, i.e., exchange the entire string. Applying the exchange function on the entire cell value can capture misaligned attributes or errors due to completely erroneous values. However, the same transformed string can also be obtained by applying either the exchange character function $T:12\longmapsto 152$ on the `12' substring of ``60612'' or the add character function $T:\varnothing \longmapsto 5$, where the position between `1' and `2' in ``60612'' was chosen at random. The distribution that corresponds to the aforementioned generative process dictates the likelihood of each of the above three cases.

Given $\Phi$ and $\Pi$, we can use this noisy channel on training examples that correspond to clean tuples to augment the available training data. However, both $\Phi$ and $\Pi$ have to be learned from the limited number of training data. This is why we adopt the above simple generative process. Despite its simplicity, we find our approach to be effective during data augmentation (see Section~\ref{sec:experiments}). Next, we introduce algorithms to learn $\Phi$ and $\Pi$ assuming access to labeled pairs of correct and erroneous values $L = \{(v^*, v)\}$ with $v \neq v^*$. We then discuss how to construct $L$ either by taking a subset of the input training data $T$ or, in the case of limited training data, via an unsupervised approach over dataset $D$. Finally, we describe how to use $\Phi$ and $\Pi$ to perform data augmentation.

\subsection{Learning Transformations}
\label{sec:phi}
We use a pattern matching approach to learn the transformations $\Phi$. We follow a hierarchical pattern matching approach to identify all different transformations that are valid for each example in $L$. For example, for $(60612, 6061x2)$ we want to extract the transformations $\{60612\longmapsto 6061x2, 12 \longmapsto 1x2, \varnothing\longmapsto x\}$. The approach we follow is similar to the Ratcliff-Obershelp pattern recognition algorithm~\cite{Ratcliff}. Due to the generative model we described above, we are agnostic to the position of each transformation.

The procedure is outlined in Algorithm~\ref{alg:phi}. Given an example $(v^*,v)$ from $L$, it returns a list of valid transformations $\Phi_e$ extracted from the  example. The algorithm first extracts the string level transformation $T: v^* \longmapsto v$, and then proceeds recursively  to extract additional transformations from the substrings of $v^*$ and $v$. To form the recursion, we identify the longest common substring of $v^*$ and $v$, and use that to split each string into its prefix (denoted by $lv^*$) and its postfix (denoted by $rv^*$). Given the prefix and the postfix substrings, we recurse on the combination of substrings that have the maximum similarity (i.e., overlap). We compute the overlap of two strings as $2*C/S$, where $C$ is the number of common characters in the two strings, and $S$ is the sum of their lengths. Finally, we remove all identity (i.e., trivial) transformations from the output $\Phi_e$. To construct the set of transformations $\Phi$, we take the set-union of all lists $\Phi_e$ generated by applying Algorithm~\ref{alg:phi} to each entry $e \in L$.

\begin{algorithm}
\small
\SetAlgoLined
\KwIn{Example $e = (v^*,v)$ of a correct string and its corresponding erroneous string}
\KwOut{A list of valid transformations $\Phi_e$ for example $e$}
 \textbf{if~}$v^* = \varnothing$ and $v = \varnothing$~\textbf{return} $\emptyset$ \;
 $\Phi_e \leftarrow [v^* \longmapsto v]$\;
 $l \leftarrow \text{Longest Common Substring}(v^*,v)$\;
 $lv^*,rv^* \leftarrow v^* \setminus l$ /* Generate left and right substrings */\;
 $lv,rv \leftarrow v\setminus l$\;
  \eIf{$\text{similarity}(lv^*, lv)+\text{similarity}(rv^*, rv) > \text{similarity}(lv^*, rv)+\text{similarity}(rv^*, lv)$}{
  Add $[lv^* \longmapsto lv,~rv^* \longmapsto rv]$ in $\Phi_e$\;
  Add $[\text{TL}(lv^*,lv),~\text{TL}(rv^*,rv)]$ in $\Phi_e$\;
   }{
    Add  $[lv^* \longmapsto rv, ~rv^* \longmapsto lv]$ in $\Phi_e$\;
    Add $[\text{TL}(lv^*,rv),~\text{TL}(rv^*,lv)]$ in $\Phi_e$\;
  }
  Remove all identity transformations from $\Phi_e$\;
  \textbf{return} $\Phi_e$
 \caption{Transformation Learning (TL)}
\label{alg:phi}
\end{algorithm}

\subsection{Policy Learning}
\label{sec:pol_learning}
The set of transformations $\Phi$ extracted by Algorithm~\ref{alg:phi} correspond to all possible alterations our noisy channel model can perform on a clean dataset. Transformations in $\Phi$ range from specialized transformations for specific entries (e.g., $60612\longmapsto 6061x2$) to generic transformations, such as $\varnothing \longmapsto x$, that can be applied to any position of any input. Given $\Phi$, the next step is to learn the transformation policy $\Pi$, i.e., the conditional probability distribution $\Pi(v) = P(\Phi|v)$ for any input value $v$. We next introduce an algorithm to learn $\Pi$.

\begin{algorithm}
\small
\SetAlgoLined
\KwIn{A set of identified transformation lists $\{\Phi_e\}_{e \in L}$}
\KwOut{Empirical Distribution  $\hat{\Pi}$}
$\Phi \leftarrow $ Set of unique transformations in $\{\Phi_e\}_{e \in L}$\;
$c \leftarrow \sum_{e} \left(\text{element count of }\Phi_e \right)$\;
\For{$\phi \in \Phi$ } {
 $c_{\phi} \leftarrow $ number of times $\phi$ appears in $\{\Phi_e\}_{e \in L}$\;
 $p(\phi) \leftarrow \frac{c_{\phi}}{c}$
 }
 \textbf{return} $\{p(\phi)\}_{\phi \in \Phi}$
\caption{Empirical Transformation Distribution}
\label{alg:pi1}
\end{algorithm}

We approximate $\Pi$ via a two-step process: First, we compute the empirical distribution of transformations informed by the transformation lists output by Algorithm~\ref{alg:phi}. This process is described in Algorithm~\ref{alg:pi1}. Second, given an input string $v$, we find all transformations $str \longmapsto str^{\prime}$ in $\Phi$ such that $str$ is a subset of $v$. Let $\Phi_v \subseteq \Phi$ be the set of such transformations. We obtain a distribution $P(\Phi_v|v)$ by re-normalizing the empirical probabilities from the first step. This process is outlined in Algorithm~\ref{alg:pi2}. Recall that we choose this simple model for $\Pi$ as the number of data points in $L$ can be limited.

\begin{algorithm}
\small
\SetAlgoLined
\KwIn{An empirical transformation $\hat{\Pi}$ over transformations $\Phi$; A string $v$}
\KwOut{Conditional Distribution  $\hat{\Pi}(v) = P(\Phi|v)$}
$\hat{\Pi}(v) \leftarrow \varnothing$\;
$\Phi_v \leftarrow $ Subset of transformations $str \longmapsto str^{\prime}$ in $\Phi$ such that $str$ is a substring of $v$\;
total mass $\leftarrow \sum_{\phi \in \Phi_v}\hat{\Pi}(\phi)$\;
\For{$\phi \in \Phi_v$ } {
 $\hat{\Pi}(v)[\phi] \leftarrow \frac{\hat{\Pi}(\phi)}{\text{total mass}}$\;
 }
\textbf{return} $\hat{\Pi}(v)$
\caption{Approximate Noisy Channel Policy}
\label{alg:pi2}
\end{algorithm}

\subsection{Generating Transformation Examples}
\label{sec:gen_l}

We describe how to obtain examples $(v^*, v)$ to form the set $L$, which we use in learning the transformations $\Phi$ (Section~\ref{sec:phi}) and the policy $\hat{\Pi}$ (Section~\ref{sec:pol_learning}). First, any example in the training data $T$ that corresponds to an error can be used. However, given the scarcity of errors in some datasets, examples of errors can be limited. We introduce a methodology based on weak-supervision to address this challenge. 

We propose a simple unsupervised data repairing model $M_R$ over dataset $D$ and use its predictions to obtain transformation examples $(v^*, v)$. We form examples $(v^*, v) = (\hat{v}, v)$ with $\hat{v} \neq v$ by taking an original cell value $v$ and the  repair $\hat{v}$ suggested by $M_R$. We only require that this model has relatively high-precision. High-precision implies that the repairs performed by $M_R$ are accurate, and thus, the predictions correspond to true errors. This approach enables us to obtain noisy training data that correspond to {\em good samples} from the distribution of errors in $D$. We do not require this simple prediction model to have high recall, since we are only after producing example errors, not repairing the whole data set.

We obtain a simple high-precision data repairing model by training a Na\"ive Bayes model over Dataset $D$. Specifically, we iterate over each cell in $D$, pretend that its value is missing and leverage the values of other attributes in the tuple to form a Na\"ive Bays model that we use to impute the value of the cell. The predicted value corresponds to the suggested repair for this cell. Effectively, this model takes into account value co-occurrence across attributes. Similar models have been proposed in the literature to form sets of potential repairs for noisy cells~\cite{hc}. To ensure high precision, we only accept only repairs with a likelihood more than 90\%. In Section~\ref{sec:experiments}, we evaluate our Na\"ive Bayes-based model and show that it achieves reasonable precision (i.e., above 70\%).

\subsection{Data Augmentation}
\label{sec:augment}
To perform data augmentation, we leverage the learned $\Phi$ and $\hat{\Pi}$ and use the generative model described in Section~\ref{sec:nc_model}. Our approach is outlined in Algorithm~\ref{alg:aug}: First, we sample a correct example with cell value $v$ from the training data $T$. Second, we sample a transformation $\phi$ from distribution $\hat{\Pi}[v]$. If $\phi$ can be applied in multiple positions or substrings of input $v$ we choose one uniformly at random, and finally, compute the transformed value $v^{\prime} = \phi(v)$. Value $v^{\prime}$ corresponds to an error as we do not consider the identity transformation. Finally, we add $(v, v^{\prime})$ in the set of augmented examples with probability $\alpha$. Probability $\alpha$ is a hyper-parameter of our algorithm, which intuitively corresponds to the required balance in the overall training data. We set $\alpha$ via cross-validation over a holdout-set that corresponds to a subset of $T$. This is the same holdout-set used to perform Platt scaling during error classification (see Section~\ref{sec:classification}).

\begin{algorithm}[ht]
\label{alg:dist}
\small
\SetAlgoLined
\KwIn{Training set $T$; Transformations $\Phi$; Approximate Policy $\hat{\Pi}$; Probability $\alpha$ (hyper-parameter)}
\KwOut{Set $T_H$ of augmented examples}
$T_H \leftarrow \emptyset$\;
$T_c \leftarrow$ set of correct examples in $T$\;
$p \leftarrow$ number of correct examples in $T$\;
$n \leftarrow$ number of erroneous examples in $T$\;
/* we assume that p $>>$ n due to imbalance */ \;
\While{$|T_H| < p - n$}{
Draw a correct example $ v \sim Uniform(T_c) $\;
$C \leftarrow $ Flip a coin with probability $\alpha$\;
\If{$C = \mathtt{True}$ and $\hat{\Pi}(v) \neq \emptyset $}{
Draw a transformation $\phi \sim \hat{\Pi}(v)$\;
$v^{\prime} \leftarrow \phi(v)$\;
$T_H \leftarrow T_H \cup \{(v, v^{\prime})\}$
 }
 }
 \caption{Data Augmentation}
 \label{alg:aug}
\end{algorithm}

\section{Experiments}
\label{sec:experiments}
We compare our approach against a wide-variety of error detection methods on diverse datasets. The main points we seek to validate are: (1) is weak supervision the key to high-quality (i.e., high-precision and high-recall) error detection models, (2) what is the impact of different representation contexts on error detection, (3) is data augmentation the right approach to minimizing human exhaust. We also perform extensive micro-benchmark experiments to examine the effectiveness and sensitivity of data augmentation.

\begin{table}[t]
\center
\scriptsize
\caption{Datasets used in our experiments.}
\label{tab:datasets}
\begin{tabular}{|c|c|c|c|c|}
\hline
{\bf Dataset} & {\bf Size} & {\bf Attributes} & {\bf Labeled Data} & {\bf Errors (\# of cells)}\\ \hline
Hospital & 1,000 & 19 & 1,000 & 504\\
Food & 170,945 & 15 & 3,000 & 1,208\\
Soccer & 200,000 & 10 & 200,000 & 31,296\\
Adult & 97,684 & 11 & 97,684 & 1,062\\
Animal & 60,575 & 14 & 60,575 & 8,077\\
\hline
\end{tabular}
\end{table}

\subsection{Experimental Setup}
\label{sec:exp_setup}
We describe the dataset, metrics, and settings we use.

\vspace{3pt}\noindent{\bf Datasets:} We use five datasets from a diverse array of domains. Table~\ref{tab:datasets} provides information for these datasets. As shown the datasets span different sizes and exhibit various amounts of errors: (1) The Hospital dataset is a benchmark dataset used in several data cleaning papers~\cite{holistic,hc}. Errors are artificially introduced by injecting typos. This is an easy benchmark dataset; (2) The Food dataset contains information on food establishments in Chicago. Errors correspond to conflicting zip codes for the same establishment, conflicting inspection results for the same establishment on the same day, conflicting facility types for the same establishment and many more. Ground truth was obtained by manually labeling 3,000 tuples; (3) The Soccer dataset provides information about soccer players and their teams. The dataset and its ground truth are provided by Rammerlaere and Geerts~\cite{Rammelaere:2018:ERD:3236187.3269456}; (4) Adult contains census data is a typical dataset from the UCI repository. Adult is also provided by Rammerlaere and Geerts~\cite{Rammelaere:2018:ERD:3236187.3269456}; (5) Animal was provided by scientists at UC Berkeley and has been used by Abedjan et al.~\cite{AbedjanCDFIOPST16} as a testbed for error detection. It provides information about the capture of animals, including the time and location of the capture and other information for each captured animal. The dataset comes with manually curated ground truth. The datasets used in our experiments exhibit different error distributions. Hospital contains only typos, Soccer~\cite{Rammelaere:2018:ERD:3236187.3269456} and Adult~\cite{Rammelaere:2018:ERD:3236187.3269456} have errors that were introduced with BART~\cite{Arocena2015}: Adult has 70\% typos and 30\% value swaps, and Soccer has 76\% typos and 24\% swaps. Finally, the two datasets with real-world errors have the following error distributions: Food has 24\% typos and 76\% value swaps (based on the sampled ground truth); Animal has 51\% typos and 49\% swaps.

\begin{table*}[t]
\center
\scriptsize
\caption{Precision, Recall and $F_1$-score of different methods for different datasets. AL results correspond to $k=100$.}
\label{tab:endres}
\begin{threeparttable}
\begin{tabular}{c l|c |c c c c c| c c c}
\shortstack{Dataset\\ ($T$ size)}& M & AUG & CV & HC & OD & FBI & LR & SuperL & SemiL & ActiveL \\ \hline \hline
\multirow{3}{*}{\shortstack{Hospital\\ (10\%)}} & P & 0.903 & 0.030 & 0.947 & 0.640 & 0.008 & 0.0 & 0.0 & 0.0 & 0.960\\
 & R & 0.989 & 0.372 & 0.353 & 0.667 & 0.001 & 0.0 & 0.0 & 0.0 & 0.613 \\
 & $F_1$ & \bf 0.944 & 0.055 & 0.514 & 0.653 & 0.003 & 0.0 & 0.0 & 0.0 & 0.748\\ \hline
\multirow{3}{*}{\shortstack{Food\\ (5\%)}} & P & 0.972 & 0.0 & 0.0 & 0.240 & 0.0 & 0.0 & 0.985 & 0.813 & 0.990\\
 & R & 0.939 & 0.0 & 0.0 & 0.99 & 0.0 & 0.0 & 0.95 & 0.66 & 0.91\\
 & $F_1$ & \bf 0.955 & 0.0 & 0.0 & 0.387 & 0.0 & 0.0 & 0.948 & 0.657 & 0.948\\\hline
\multirow{3}{*}{\shortstack{Soccer\\ (5\%)}} & P & 0.922 & 0.039 & 0.032 & 0.999 & 0.0 & 0.721 & 0.802 & n/a\tnote{\#} & 0.843\\
 & R & 1.0 & 0.846 & 0.632 & 0.051 &0.00 & 0.084 & 0.450 & n/a & 0.683\\
 & $F_1$ & \bf 0.959 & 0.074 & 0.061 & 0.097 &0.00 & 0.152 & 0.577 & n/a & 0.755\\ \hline 
\multirow{3}{*}{\shortstack{Adult\\ (5\%)}} & P & 0.994 & 0.497 & 0.893 & 0.999 & 0.990 & 0.051 & 0.999 & n/a & 0.994\\
 & R & 0.987 & 0.998 & 0.392 & 0.001 & 0.254 & 0.072 & 0.350 & n/a& 0.982\\
 & $F_1$ & \bf 0.991 & 0.664 & 0.545 & 0.002 & 0.405 & 0.059 & 0.519 & n/a & 0.988\\ \hline
 \multirow{3}{*}{\shortstack{Animal\\ (5\%)}} & P & 0.832 & 0.0 & 0.0 & 0.85 & 0.0 & 0.185 & 0.919 & n/a & 0.832\\
 & R & 0.913 & 0.0 & 0.0 & $6\times10^{-5}$ & 0.0 & 0.028 & 0.231 & n/a & 0.740\\
 & $F_1$ & \bf 0.871 & 0.0 & 0.0 & $1\times10^{-4}$ & 0.0 & 0.048 & 0.369 & n/a &0.783\\ \hline \\
\end{tabular}
\begin{tablenotes}
  \item[\#] n/a = Semi-supervised learning did not terminate after two days.
  \end{tablenotes}
\end{threeparttable}
\end{table*}

\vspace{3pt}\noindent{\bf Methods:} We compare our approach, referred to as \texttt{AUG}, against several competing error detection methods. First, we consider three baseline error detection models:
\begin{itemize}
	\item \textbf{Constraint Violations (CV)}: This method identifies errors by leveraging violations of denial constraints. It is a proxy for rule-based errors detection methods~\cite{holistic}.
	\item \textbf{HoloClean (HC)}: This method combines CV with HoloClean~\cite{hc}, a state-of-the-art data repairing engine. This method aims to improve the precision of the CV detector by considering as errors not all cells in tuples that participate in constraint violations but only those cells whose value was repaired (i.e., their initial value is changed to a different value).
	\item \textbf{Outlier Detection (OD)}: This method follows a correlation based outlier detection approach. Given a cell that corresponds to an attribute $A_i$, the method considers all correlated attributes in $A \setminus A_i$ with $A_i$ rely on the pair-wise conditional distributions to detect if the value of a cell corresponds to an outlier.
	\item \textbf{Forbidden Item Sets (FBI)}: This method captures unlikely value co-occurrences in noisy data~\cite{rammelaere2017cleaning}. At its core, this method leverages the {\em lift} measure from association rule mining to identify how probably a value co-occurrence is, and uses this measure to identify erroneous cell values.
	\item \textbf{Logistic Regression (LR)}: This method corresponds to a supervised logistic regression model that classifies cells are erroneous or correct. The features of this model correspond to pairwise co-occurrence statistics of attribute values and constraint violations. This model corresponds to a simple supervised ensemble over the previous two models.
\end{itemize}
We also consider three variants of our model where we use different training paradigms. The goal is to compare data augmentation against other types of training. For all variations, we use the representation $Q$ and the classifier $M$ introduced in Section~\ref{sec:overview}. We consider the following variants:
\begin{itemize}
	\item \textbf{Supervised Learning (SuperL)}: We train our model using only the training examples in $T$.
	\item \textbf{Semi-supervised Learning (SemiL)}: We train our model using self-training~\cite{zhu2007semi}. First supervised learning used to train the model on the labeled data only. The learned model is then applied to the entire dataset to generate more labeled examples as input for a subsequent round of supervised learning. Only labels with high confidence are added at each step.
	\item \textbf{Active Learning (ActiveL)}: We train our model using an active learning method based on uncertainty sampling~\cite{settles2012active}. First, supervised learning is used to train the model. At each subsequent round, we use an uncertainty-based selection scheme to obtain additional training examples and re-train the model. We use $k$ to denote the number of iterations. In our implementation, we set the upper limit of labeled examples obtained per iteration to be $50$ cells.
\end{itemize}

\vspace{3pt}\noindent{\bf Evaluation Setup:} To measure accuracy, we use Precision (P) defined as the fraction of error predictions that are correct; Recall (R) defined as the fraction of true error being predicted as errors by the different methods; and $F_1$ 
defined as $2PR/(P + R)$. For training, we split the available ground truth into three disjoint sets: (1)~a training set $T$, from which $10\%$ is always kept as a hold-out set used for hyper parameter tuning; (2)~a sampling set, which is used to obtain additional labels for active learning;  and (3)~a test set, which is used for evaluation. To evaluate different dataset splits, {\em we perform $10$ runs with different random seeds for each experiment}. To ensure that we maintain the coupling amongst Precision, Recall, and $F_1$, we report the median performance. The mean performance along with standard error measurements are reported in the Appendix. {\em Seeds are sampled at the beginning of each experiment, and hence, a different set of random seeds can be used for different experiments.} We use {\it ADAM}~\cite{journals/corr/KingmaB14} as the optimization algorithm for all learning-based model and train all models for 500 epochs with a batch-size of five examples. We run Platt Scaling for 100 epochs. All experiments were executed on a 12-core Intel(R) Xeon(R) CPU E5-2603 v3 @ 1.60GHz with 64GB of RAM running Ubuntu 14.04.3 LTS.

\subsection{End-to-end Performance}
\label{sec:end_to_end}

\begin{figure*}
  \centering
  \includegraphics[width=0.8\textwidth]{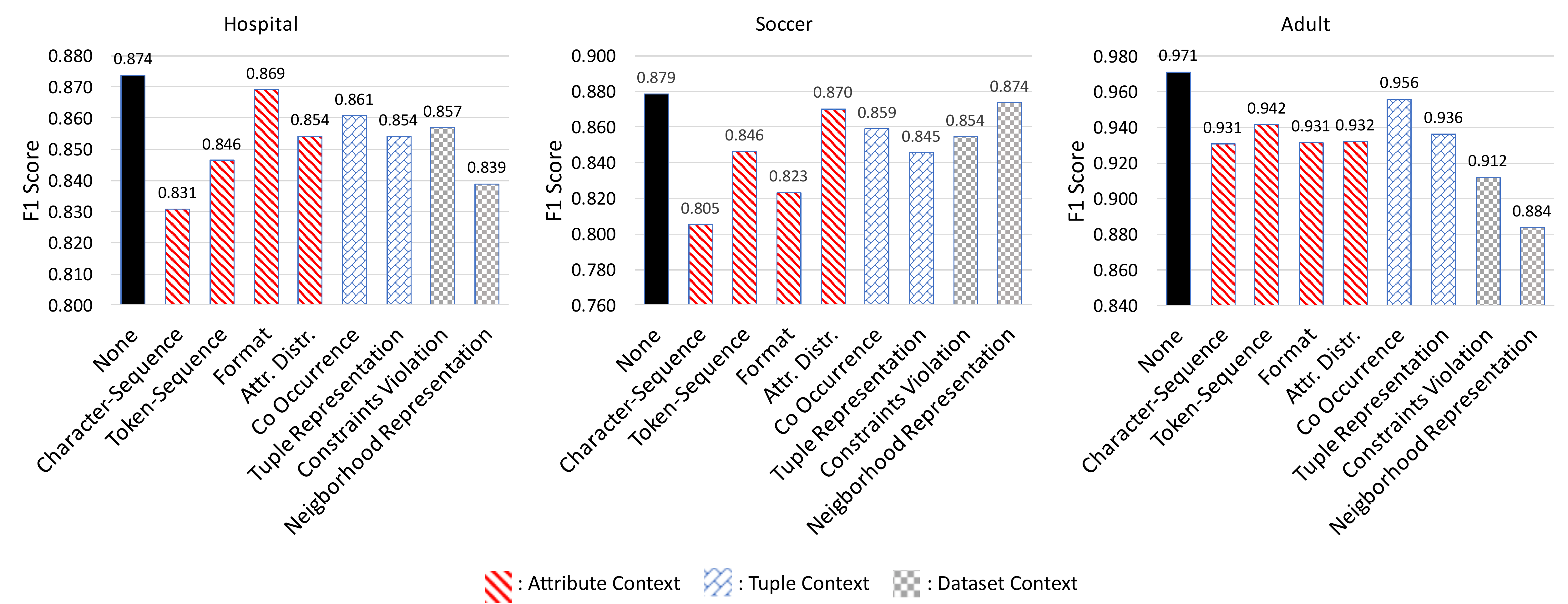}
  \caption{Ablation studies to evaluate the effect of different representation models.}
  \label{fig:ablation}
\end{figure*}

We evaluate the performance of our approach and competing approaches on detecting errors in all five datasets. Table~\ref{tab:endres} summarizes the precision, recall, and $F_1$-score obtained by different methods. For Food, Soccer, Adult, and Animal we set the amount of training data to be $5\%$ of the total dataset. For Hospital we set the percentage of training data to be $10\%$ (corresponding to 100 tuples) since Hospital is small. For Active Learning we set the number of active learning loops to $k=100$ to maximize performance.

As Table~\ref{tab:endres} shows, our method consistently outperforms all methods, and in some cases, like Hospital and Soccer, we see improvements of 20 $F_1$ points. More importantly, we find that our method is able to achieve both high recall and high precision in all datasets despite the different error distribution in each dataset. This is something that has been particularly challenging for prior error detection methods. We see that for Food and Animal, despite the fact that most errors do not correspond to constraint violations (as implied by the performance of CV), AUG can obtain high precision and recall. This is because AUG models the actual data distribution and not the side-effects of errors. For instance, for Food we see that OD can detect many of the errors---it has high recall---indicating that most errors correspond to statistical outliers. We see that AUG can successfully solve error detection for this dataset. Overall, our method achieves an average precision of $92\%$ and an average recall of $96\%$ across these diverse datasets. At the same time, we see that the performance of competing methods varies significantly across datasets. This validates the findings of prior work~\cite{AbedjanCDFIOPST16} that depending on the side effects of errors different error detection methods are more suitable for different datasets.  

We now discuss the performance of individual competing methods. For CV, we see that it achieves higher recall than precision. This performance is due to the fact that CV marks as erroneous all cells in a group of cells that participate in a violation. More emphasis should be put on the recall-related results of CV. As shown its recall varies dramatically from 0.0 for Food and Animal to 0.998 for Adult. For OD, we see that it achieves relatively high-precision results, but its recall is low. Similar performance is exhibited by FBI that leverages a different measure for outlier detection. We see that FBI achieves high precision when the forbidden item sets have significant support (i.e., occur relatively often). However, FBI cannot detect errors that lead to outlier values which occur a limited number of times. This is why we find OD to outperform FBI in several cases.

Using HC as a detection tool is limited to these cells violating integrity constraints. Hence, using HC leads to improved precision over CV (see Hospital and Adult). This result is expected as data repairing limits the number of cells detected as erroneous to only those whose values are altered. Our results also validate the fact that HC depends heavily on the quality of the error detection used~\cite{hc}. As shown in Food and Animal, the performance of HC is limited by the recall of CV, i.e., since CV did not detect errors accurately, HC does not have the necessary training data to learn how to repair cells. At the same time, Soccer reveals that training HC on few clean cells---the recall of CV is very high while the precision is very low indicating that most cells were marked as erroneous---leads to low precision (HC achieves a precision of 0.032 for Soccer). This validates our approach of solving error detection separately from data repairing.


We also see that LR has consistently poor performance. This result reveals that combining co-occurrence features and violations features in a linear way (i.e., via a weighted linear combination such as in LR) is not enough to capture the complex statistics of the dataset. {\em This validates our choice of using representation learning and not engineered features}.

Finally, we see that approaches that rely on representation learning model achieve consistently high precision across all datasets. This validates our hypothesis that modeling the distribution of both correct and erroneous data allows us to discriminate better. However, we see that when we rely only on the training dataset $T$ the recall is limited (see the recall for SuperL). The limited labeled examples in $T$ is not sufficient to capture the heterogeneity of errors. Given additional training examples either via Active Learning or via Data Augmentation helps improve the recall. However, Data Augmentation is more effective than Active Learning at capturing the heterogeneity of errors in each dataset, and hence, achieves superior recall to Active Learning in all cases.

\noindent\textbf{Takeaway:} The combination of representation learning techniques with data augmentation is key to obtaining high-quality error detection models.

\begin{figure*}
  \centering
  \includegraphics[width=0.7\textwidth]{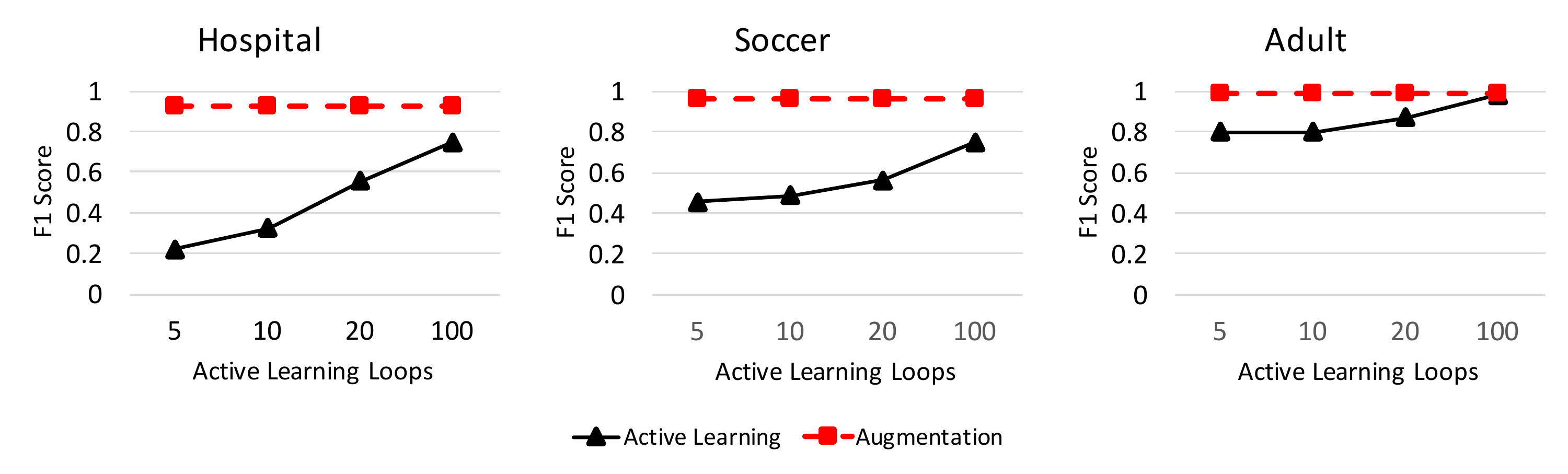}
  \caption{Data augmentation versus active learning as the number of active learning loops increases.}
  \label{fig:exp2}
\end{figure*}

\subsection{Representation Ablation Study}
\label{sec:ablation}

We perform an ablation study to evaluate the effect of different representation models on the quality of our model. Specifically, we compare the performance of AUG when all representation models are used in $Q$ versus variants of AUG where one model is removed at a time. We report the $F_1$-score of the different variants as well as the original AUG in Figure~\ref{fig:ablation}. Representation models that correspond to different contexts are grouped together.

Removing any feature has an impact on the quality of predictions of our model. We find that removing a single representation model results in drops of up to 9 $F_1$ points across datasets. More importantly, we find that different representation models have different impact on different datasets. For instance, the biggest drop for Hospital and Soccer is achieved when the character-sequence model is removed while for Adult the highest drop is achieved when the Neighborhood representation is removed. This validates our design of considering representation models from different contexts.
\noindent\textbf{Takeaway:} It is necessary to leverage cell representations that are informed by different contexts to provide robust and high-quality error detection solutions.
\subsection{Augmentation versus Active Learning}
\label{sec:aug_vs_al}
We validate the hypothesis that data augmentation is more effective than active learning in minimizing human effort in training error detection models. In Table~\ref{tab:endres}, we showed that data augmentation outperforms active learning. Furthermore, active learning needs to obtain more labeled examples to achieve comparable performance to data augmentation. In the next two experiments, we examine the performance of the two approach as we limit their access to training data. 

In the first experiment, we evaluate active learning for different values of loops ($k$) over Hospital, Soccer, and Adult. We vary $k$ in $\{5,10,20,100\}$. We fix the amount of available training data to $5\%$. Each time we measure the $F_1$ score of the two algorithms. We report our results in Figure~\ref{fig:exp2}. Reported results correspond to median performance over ten runs.

We see that when a small number of loops is used (k=$5$), there is a significant gap between the two algorithms that ranges between $10$ and $70$ $F_1$ points. Active learning achieves comparable performance with data augmentation only after $100$ loops. This corresponds to an additional $5,000$ ($k \times 50$) labeled examples (labeled cells). This behavior is consistent across all three datasets. 
\begin{figure}
  \centering
  \includegraphics[width=0.32\textwidth]{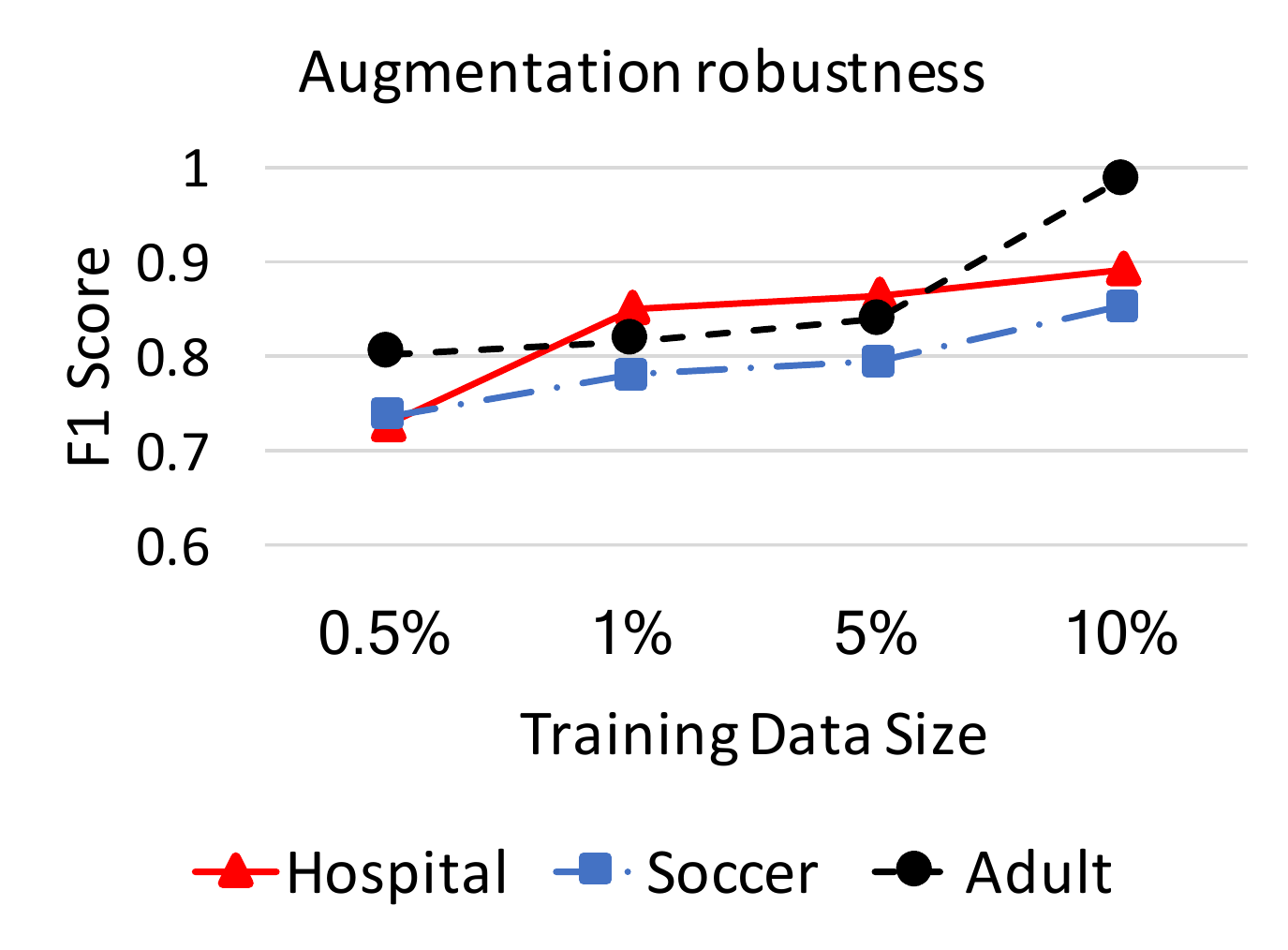}
  \caption{Data augmentation performance for various amounts of training data.}
  \label{fig:exp7}
\end{figure}
In the second experiment, we seek to push data augmentation to the limits. Specifically, we seek to answer the question, can data augmentation be effective when the number of labeled examples in $T$ is extremely small. To this end, we evaluate the performance of our system on Hospital, Soccer, and Adult as we vary the size of the training data in $\{0.5\%, 1\%, 5\%, 10\%\}$. The results are shown in Figure~\ref{fig:exp7}. As expected the performance of data augmentation is improving as more training data become available. However, we see that data augmentation can achieve good performance---$F_1$ score does not drop below 70\%---even in cases where labeled examples $T$ are limited. These results provide positive evidence that data augmentation is a viable approach for minimizing user exhaust.

\noindent\textbf{Takeaway:} Our data augmentation approach is preferable to active learning for minimizing human exhaust.

\begin{figure*}
  \centering
  \includegraphics[width=0.7\textwidth]{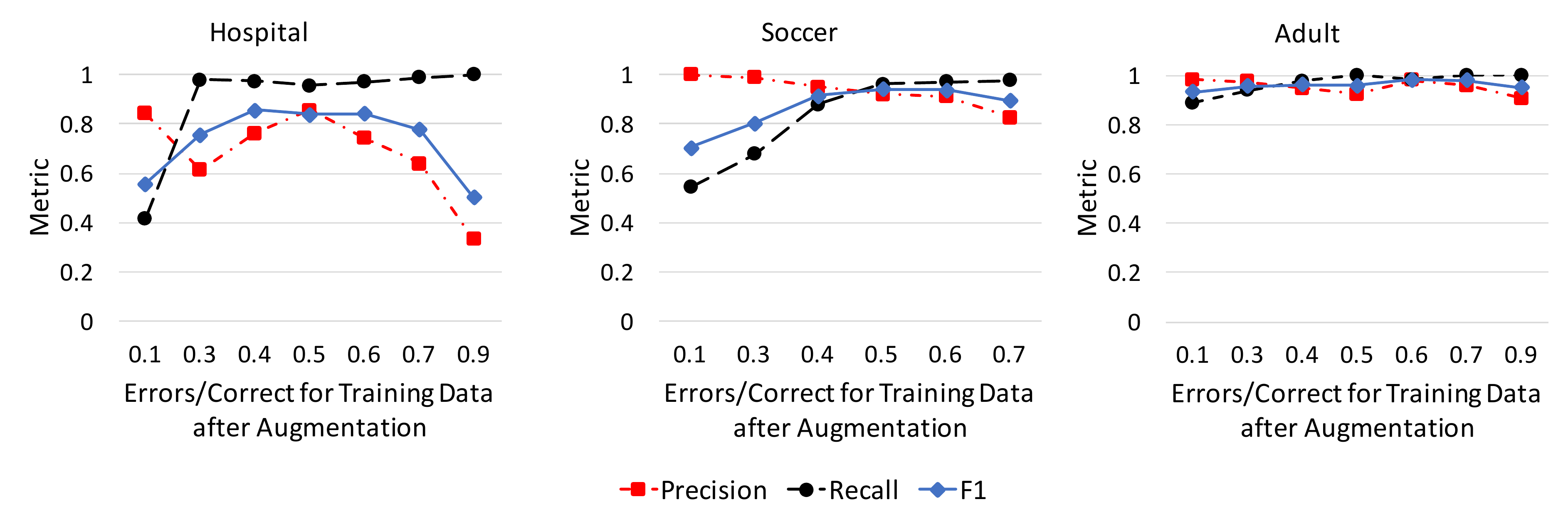}
  \caption{The effect of increasing the number of examples that correspond to errors via data augmentation.}
  \label{fig:exp3}
\end{figure*}

\subsection{Augmentation and Data Imbalance}
\label{sec:data_imbalance}
\begin{table}[t]
\center
\small
\caption{A comparison between data augmentation and resampling. We report the $F_1$-score as we increase the size of the training data $T$. We also include supervised learning as a baseline.}
\label{tab:rebalance}
\begin{tabular}{c l|c c c}
Dataset & Size of $T$ & AUG & Resampling & SuperL\\ \hline \hline
\multirow{3}{*}{Hospital} & 1\% & \bf 0.840 & 0.041 & 0.0\\
 & 5\% & \bf 0.873 & 0.278 & 0.0\\
 & 10\% & \bf 0.925 & 0.476 & 0.079\\ \hline
\multirow{3}{*}{Soccer} & 1\% & \bf 0.927 & 0.125 & 0.577\\
 & 5\% & \bf 0.935 & 0.208 & 0.654\\
 & 10\% & \bf 0.953 & 0.361 & 0.675\\ \hline
 \multirow{3}{*}{Adult} & 1\% & \bf 0.844 & 0.063 & 0.0\\
 & 5\% & \bf 0.953 & 0.068 & 0.294\\
 & 10\% & \bf 0.975 & 0.132 & 0.519\\ \hline
\end{tabular}
\end{table}
We evaluate the effectiveness of data augmentation to counteract imbalance. Table~\ref{tab:endres} shows that using data augmentation yields high-quality error detection models for datasets with varying percentages of errors. Hence, data augmentation is robust to different levels of imbalance; each dataset in Table~\ref{tab:endres} has a different ratio of true errors to correct cells.

In Table~\ref{tab:rebalance}, we compare data augmentation with traditional methods used to solve the imbalance problem, namely, resampling. In all the datasets, resampling had low precision and recall confirming our hypothesis discussed in Section~\ref{sec:introduction}: due to the heterogeneity of the errors, resampling from the limited number of negative examples was not enough to cover all types of errors. The best result for resampling was obtained in the Hospital data set ($F_1$ about $47\%$), since errors are more homogeneous than other data sets. 

We also evaluate the effect of excessive data augmentation: In Algorithm~\ref{alg:aug} we do not use hyper-parameter $\alpha$ to control how many artificial examples should be generated via data augmentation. We manually set the ratio between positive and negative examples in the final training examples and use augmentation to materialize this ratio.

Our results are reported in Figure~\ref{fig:exp3}. We show that increasing the number of generated negative examples (errors) results in a lower accuracy as the balance between errors and correct example goes greater than $50\%$, as the model suffers from the imbalance problem again, this time as too few correct examples. We see that peak performance is achieved when the training data is almost balanced for all datasets. This reveals the robustness of our approach. Nonetheless, peak performance is not achieved exactly at a 50-50 balance (peak performance for Adult is at 60\%). This justifies our model for data augmentation presented in Algorithm~\ref{alg:aug} and the use of hyper-parameter $\alpha$.

\noindent\textbf{Takeaway:} Data augmentation is an effective way to counteract imbalance in error detection.

\subsection{Analysis of Augmentation Learning}
\label{sec:aug_learning}

In this experiment, we validate the importance of learning the augmentation model (the transformations $\Phi$, and the policy $\hat{\Pi}$). We compare three augmentation strategies: (1)~Random transformations {\it Rand. Trans.}, where we randomly choose from a set of errors (e.g., typos, attribute value changes, attribute shifts, etc.). Here, we augment the data by using completely random transformations not inspired by the erroneous examples or the data; and (2)~learned transformation $\Phi$, but  without learning the distribution policy({\it Aug w/o Policy}). Given an input, we find all valid transformations in $\Phi$ and pick one uniformly at random. Table~\ref{tab:augpolicy} shows the results for the three approaches. AUG outperforms the other two strategies. Rand. Trans. fails to capture the errors that exist in the dataset. For instance, it obtains a recall of 16.6\% for Soccer. Even though the transformations are learned from the data, it is the results show that using these transformations in a way that conform with the distribution of the data is crucial in learning an accurate classifier.

\noindent\textbf{Takeaway:} Learning a noisy channel model from the data, i.e., a set of transformations $\Phi$ and a policy $\Pi$ is key to obtaining high-quality predictions.

\begin{table}[t]
\center
\small
\caption{A comparison between different data augmentation approaches. We report the $F_1$-score as we increase the size of the training data $T$.}
\label{tab:augpolicy}
\begin{tabular}{c l|c c c}{}
Dataset & $T$ & AUG & Rand. Trans. & AUG w/o Policy\\ \hline \hline
\multirow{3}{*}{Hospital} & 5\% & \bf 0.911 & 0.873 & 0.866\\
 & 10\% & \bf 0.943 & 0.884 & 0.870\\ \hline
\multirow{3}{*}{Soccer} & 5\% & \bf 0.946 & 0.212 & 0.517\\
 & 10\% & \bf 0.953 & 0.166 & 0.522\\ \hline
 \multirow{3}{*}{Adult} & 5\% & \bf 0.977 & 0.789 & 0.754\\
 & 10\% & \bf 0.984 & 0.817 & 0.747\\ \hline
\end{tabular}
\end{table}

\subsection{Other Experiments}
\label{sec:other_exps}
Finally, we report several benchmarking results: (1) we measure the runtime of different methods, (2) validate the performance of our unsupervised Na\"ive Bayes model for generating labeled example to learn transformations $\Phi$ and $\Pi$ (see Section~\ref{sec:augment}), and (3) validate the robustness of AUG to misspecified denial constraints.

\begin{table}
\center
\small
\caption{Runtimes in seconds. Value n/a means that the method did not terminate after running for two days.}
	\begin{tabular}{c| c c c}
		\textbf{Approach}         & \textbf{Hospital} & \textbf{Soccer} & \textbf{Adult}  \\ \hline
AUG      & 749.17            & 7684.72         & 6332.13         \\ \hline
CV                & 204.62            & 1610.02         & 1359.46         \\ \hline
OD & 212.7             & 1588.06         & 1423.69         \\ \hline
LR     & 347.95            & 3505.60         & 4408.27         \\ \hline
SuperL        & 648.34            & 3928.46         & 3310.71         \\ \hline
SemiL  & 14985.15          & n/a & n/a \\ \hline
ActiveL   & 3836.15           & 56535.19        & 128132.56       \\ \hline
	\end{tabular}
\label{tab:runtimes}
\end{table}

The median runtime of different methods is reported in Table~\ref{tab:runtimes}. These runtimes correspond to prototype implementations of the different methods in Python. Also recall, that training corresponds to 500 epochs with low batch-size as reported in Section~\ref{sec:exp_setup}. As expected iterative methods such as SemiL and ActiveL are significantly slower than non-iterative ones. Overall, we see that AUG exhibits runtimes that are of the same order of magnitude as supervised methods.

\begin{table}
    \center
    \small
    \caption{Performance of our weak supervision method for generating training examples for AUG.}
	\begin{tabular}{c|cc}
\textbf{Dataset}  & \textbf{Precision} & \textbf{Recall} \\ \hline
Hospital & 0.895              & 0.636           \\ \hline
Soccer   & 0.999              & 0.053           \\ \hline
Adult    & 0.714              & 0.973           \\ \hline
\end{tabular}
\label{tab:nb_perf}
\end{table}

The performance of our Na\"ive Bayes-based weak supervision method on Hospital, Soccer, and Adult is reported in Table~\ref{tab:nb_perf}. Specifically, we seek to validate that the precision of our weak supervision method is reasonable, and thus, by using it we obtain good examples that correspond to good examples from the true error distribution. We see that our weak supervision method achieves a precision of more than 70\% in all cases. As expected its recall can be some times low (e.g., for Soccer it is 5.3\%) as emphasis is put on precision.

Finally, we evaluate AUG against missing and noisy constraints. The detailed results are presented in Appendix~\ref{sec:more_micro} due to space restrictions. In summary, we find AUG to exhibit a drop of at most 6 $F_1$ points when only 20\% of the original constraints are used to missing constraints and at most 8 $F_1$ points when noisy constraints are used.
\section{Related Work}
\label{sec:related}

Many algorithms and prototypes have been proposed for developing data cleaning tools~\cite{Rahm00,IlyasC15,Fan2012,HalevyBook}. Outlier detection and quantitative data cleaning algorithms are after data values that looks ``abnormal'' with respect to the data distribution~\cite{Dasu2012,Wu:2013,2015combining}. Entity resolution and record de-duplication focus on identifying clusters of records that represent the same real-world entity~\cite{Elmagarmid07,2010Naumann}. Example de-duplication tools include the {\tt Data Tamer} system~\cite{Stonebraker13}, which is commercialized as {\tt Tamr}. Rule-based detection proposals ~\cite{abedjan_2015,holistic,WangT14,FanLMTY12,Kolahi09} use integrity constraints (e.g., denial constraints) to identify violations, and use the overlap among these violations to detect data errors. Prototypes such as such as {\tt Nadeef}~\cite{dallachiesa2013nadeef}, and {\tt BigDansing}~\cite{Khayyat:2015} are example extensible rule-based cleaning systems. There have been also multiple proposals that identify data cells that don not follow a data ``pattern''. Example tools include {\tt OpenRefine}, {\tt Data Wrangler}~\cite{wrangler} and its commercial descendant {\tt Trifacta}, {\tt Katara}~\cite{Chu15}, and {\tt DataXFormer}~\cite{Abedjan_2016}. An overview  of these tools and how they can be combined for error detection is discussed in~\cite{AbedjanCDFIOPST16}, where the authors show that even when all are used, these tools often achieve low recall in capturing data errors in real data sets.

Data Augmentation has also been used extensively in machine learning problems. Most state-of-the-art image classification pipelines use some limited for of data augmentation~\cite{Perez2017TheEO}. This consists of applying crops, flips, or small affine transformations in fixed order or at random. Other studies have applied heuristic data augmentation to modalities such as audio~\cite{sound_da} and text~\cite{lu2006enhancing}. To our knowledge, we are the first to apply data augmentation in relational data.

Recently, several lines of work have explored the use of reinforcement learning or random search to learn more principled data augmentation policies~\cite{cubuk2018autoaugment,DBLP:conf/nips/RatnerEHDR17}.  Our work here is different as we do not rely on expensive procedures to learn the augmentation policies. This is because we limit our policies to applying a single transformation at a time. Finally, recent work has explored techniques based on Generative Adversarial Networks~\cite{NIPS2014_5423} to learn data generation models used for data augmentation from unlabeled data~\cite{Mirza2014ConditionalGA}. This work focuses mostly on image data. Exploring this direction for relational data is an exciting future direction.

\section{Conclusions}
\label{sec:conclusions}

We introduced a few-shot learning error detection framework. We adopt a noisy channel model to capture how both correct data and errors are generated use it to develop an expressive classifier that can predict, with high accuracy, whether a cell in the data is an error. To capture the heterogeneity of data distributions, we learn a rich set of representations at various granularities (attribute-level, record-level, and the dataset-level). We also showed how to address a main hurdle in this approach, which is the scarcity of error examples in the training data, and we introduced an approach based on data augmentation to generate enough examples of data errors. Our data augmentation approach learns a set of transformations and the probability distribution over these transformations from a small set of examples (or in a completely unsupervised way). We showed that our approach achieved an average precision of \textasciitilde94\% and an average recall of \textasciitilde93\% across a diverse array of datasets. We also showed how our approach outperforms previous techniques ranging from traditional rule-based methods to more complex ML-based method such as active learning approaches. 

\section{Acknowledgements} This work was supported by Amazon under an ARA Award, by NSERC under a Discovery Grant, and by NSF under grant IIS-1755676.

\appendix
\section{Appendix}
\label{sec:appdx}
We provide additional details for the representation models in our framework and present additional micro-benchmark experimental results on the robustness of our error detection approach to noisy denial constraints.

\subsection{Details on Representation Models}
\label{sec:model_details}
\begin{figure}
  \includegraphics[width=0.45\textwidth]{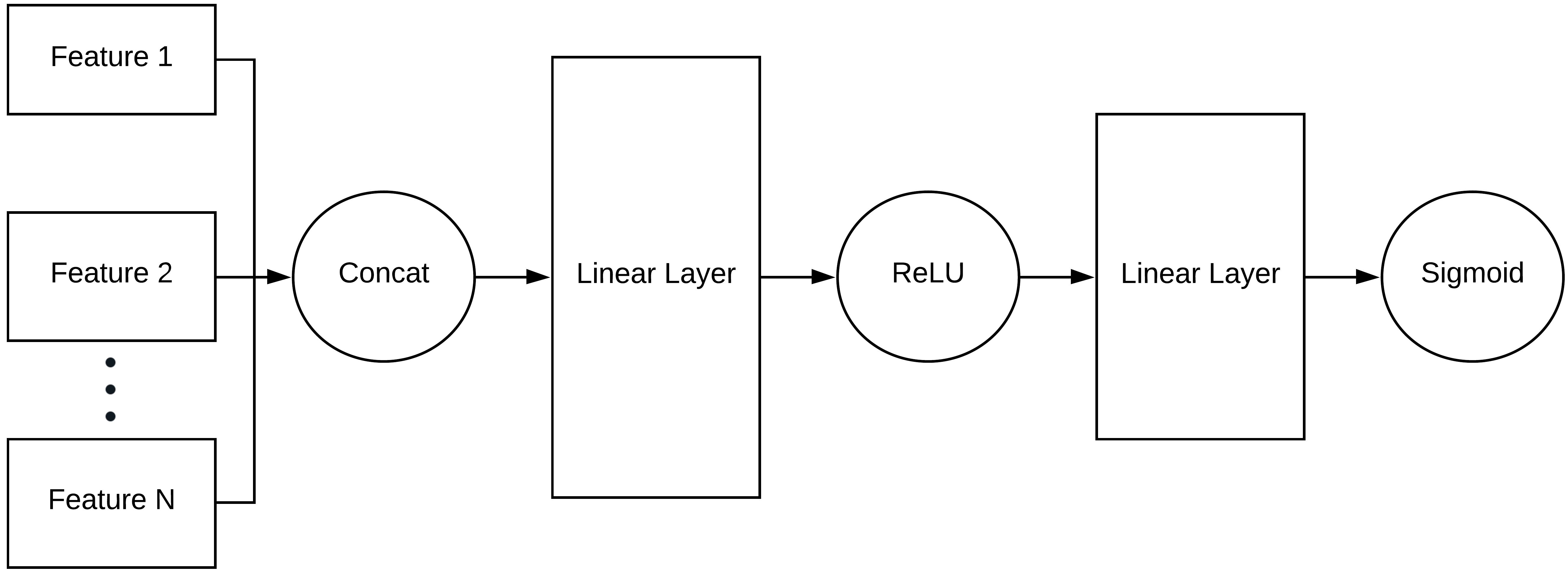}
  \caption{The architecture of our representation learning model following a wide and deep architecture.}
  \label{fig:wideanddeep}
\end{figure}

Our model follows the wide and deep architecture of Cheng et al.~\cite{wideanddeep}. Thus the model can be thought of as a representation stage, where each feature is being operated on in isolation, and an inference step in which each feature has been concatenated to make a joint representation. The joint representation is then fed through a two-layer neural network. At training time, we backpropogate through the entire network jointly, rather than training specific representations. Figure~\ref{fig:wideanddeep} illustrates this model's topology.

A summary of representation models used in our approach along with their dimensions is provided in Table~\ref{tab:modelfeatures}. As shown we use a variety of models that capture all three attribute-level, tuple-level, and dataset-level contexts. We next discuss the embedding-based models and format models we use.

\begin{table*}[t]
\center
\scriptsize
	\caption{A summary of representation models used in our approach along with their dimension.}
\label{tab:modelfeatures}
\begin{threeparttable}
	\begin{tabular}{c|c|c|c}
		Context & Representation Type & Description & Dimension \\ \hline \hline
				\multirow{ 6}{*}{Attribute-Level} & Character Embedding & FastText Embedding where tokens are characters & 1 \\
		& Word Embedding & FastText Embedding where tokens are words in the cell & 1 \\				
		& Format models & 3-Gram: Frequency of the least frequent 3-gram in the cell & 1 \\
		& Format models & Symbolic 3-Gram; each character is replaced by a token $\{Char, Num, Sym\}$& 1 \\
		& Empirical distribution model & Frequency of cell value & 1 \\
		& Empirical distribution model & One Hot Column ID; Captures per-column bias & 1 \\\hline
		\multirow{ 2}{*}{Tuple-Level} & Co-occurrence model & Co-occurrence statistics for a cell's value & \#attributes -1 \\
		& Tuple representation & FasText-based embedding of the union of tokens after tokenizing each attribute value & 1 \\\hline
				\multirow{ 2}{*}{Dataset-Level} & Constraint violations & Number of violations per denial constraint & \#constraints \\		 
		 & Neighborhood representation & Distance to top-1 similar word using a FastText tuple embedding over the non-tokenized attribute values & 1 \\\hline
\end{tabular}
\end{threeparttable}
\end{table*}

\vspace{3pt}\noindent\textbf{Embedding-based Models:} We treat different views of the data as expressing different language models, and so embed each to capture their semantics. The embeddings are taken at a character, cell 
and tuple level tokens, and each uses a FastText Embedding in 50 dimensions~\cite{E17-2068, Q17-1010}. Rather than doing inference directly on the embeddings, we employ a two-step process of a non-linear transformation and dimensionality reduction. At the non-linear transformation stage, we use a two-layer Highway Network~\cite{srivastava2015highway} to extract useful representations of the data. Then, a dense layer is used to reduce the dimensionality to a single dimension. In this way, the embeddings do not dominate the joint representation. Figure~\ref{fig:models_overview}(B) shows this module more explicitly.

In addition to using these singular embeddings, we also use a distance metric on the learned corpus as a signal to be fed into the model (see Neighborhood representation). The intuition behind this representation is that in the presence of other signals that would imply a cell is erroneous, there may be some similar cell in the dataset with the correct value; hence, the distance to it will be low. For this, we simply take the minimum distance to another embedding in our corpus, and this distance is fed to the joint representation.

\vspace{3pt}\noindent\textbf{Forma Models (3-Grams):} We follow a similar approach to that of Huang and He~\cite{DBLP:conf/sigmod/HuangH18}. This work introduces custom language models to do outlier detection. We follow a simplified variation of this approach and use two fixed length language models. They correspond to the 3-Gram models shown in Table~\ref{tab:modelfeatures}. To build these representation models, we build a distribution of 3-Grams present in each column, this is done using the empirical distribution of the data and Laplace smoothing. For 3-Gram, the distribution is based on all possible ASCII 3-Grams. The difference in the symbol based variation of 3-Gram is that the distribution is based off the alphabet $\{Charcater, Number, Symbol\}$. The value returned for each model is the least frequency of all 3-grams present in the cell value.

\subsection{Effect of Misspecified Constraints}
\label{sec:more_micro}
We conduct a series of micro-benchmark experiments to evaluate the robustness of AUG against misspecified denial constraints. First, we evaluate AUG's performance as only a subset of constraints is given as input, and second, we evaluate AUG's performance as constraints become noisy. 

\subsubsection{Limiting the number of Constraints}
\label{sec:limitingdcs}

We consider Hospital, Adult, and Soccer with the denial constraints used for our experiments in Section~\ref{sec:experiments} and perform the following experiment: For each dataset, we define a vary the number of constraints given as input to AUG by taking only a proportion $\rho$ of the initial constraints. We vary $\rho$ in $\{0.2, 0.4, 0.6$, $0.8, 1.0\}$ where $0.2$ indicates that a random subset of 20\% of the constraints is used while $1.0$ indicates that all constraints are used. For each configuration for $\rho$ we obtain 21 samples of the constraints and evaluate AUG for these random subsets. We report the median $F_1$, precision, and recall in Table~\ref{tab:lessnumberofrules}. As shown, AUGs performance gradually decreases as the number of denial constraints is reduced and converges to the performance reported in the study in Section~\ref{sec:ablation} when no constraints are used in AUG. The results in Table~\ref{tab:lessnumberofrules} also show that AUG is robust to small variations in the number of constraints provided as input. We see that when $\rho > 0.4$ the $F_1$ score of AUG does not reduce more than two points.

\begin{table}[h]
\center
\scriptsize
\caption{Median performance of AUG over 21 runs as we randomly limit the input constraints to $\rho \times |$initial constraints$|$.}
\label{tab:lessnumberofrules}
\begin{tabular}{cc|ccccc}
\textbf{Dataset}  & \textbf{M}                                        & \textbf{$ \rho = 0.2 $}                                                                    & \textbf{$ \rho = 0.4 $}                                                                    & \textbf{$ \rho = 0.6 $}                                                                    & \textbf{$ \rho = 0.8 $}                                                                    & \textbf{$ \rho = 1 $}                                                                      \\ \hline
\textbf{Hospital} & \begin{tabular}[c]{@{}c@{}}P\\ R\\ $F_1$\end{tabular} & \begin{tabular}[c]{@{}c@{}}0.857\\ 0.848\\ 0.852\end{tabular} & \begin{tabular}[c]{@{}c@{}}0.829\\ 0.877\\ 0.852\end{tabular} & \begin{tabular}[c]{@{}c@{}}0.927\\ 0.857\\ 0.891\end{tabular} & \begin{tabular}[c]{@{}c@{}}0.925\\ 0.896\\ 0.910\end{tabular} & \begin{tabular}[c]{@{}c@{}}0.936\\ 0.901\\ 0.918\end{tabular} \\ \hline
\textbf{Adult}    & \begin{tabular}[c]{@{}c@{}}P\\ R\\ $F_1$\end{tabular} &
\begin{tabular}[c]{@{}c@{}}0.860\\ 0.994\\ 0.922\end{tabular}    &
\begin{tabular}[c]{@{}c@{}}0.890\\ 0.992\\ 0.938\end{tabular}    &
\begin{tabular}[c]{@{}c@{}}0.897\\ 0.999\\ 0.945\end{tabular}    & 
\begin{tabular}[c]{@{}c@{}}0.917\\ 0.999\\ 0.956 \end{tabular}    & 
\begin{tabular}[c]{@{}c@{}}0.934\\ 0.999\\ 0.965\end{tabular}      \\ \hline
\textbf{Soccer}   & \begin{tabular}[c]{@{}c@{}}P\\ R\\ $F_1$\end{tabular} &
\begin{tabular}[c]{@{}c@{}}0.836\\ 0.868\\ 0.852\end{tabular}   &
\begin{tabular}[c]{@{}c@{}}0.855\\ 0.879\\ 0.867\end{tabular}    &  
\begin{tabular}[c]{@{}c@{}}0.864\\ 0.872\\ 0.868\end{tabular}    & 
\begin{tabular}[c]{@{}c@{}}0.860\\ 0.887\\ 0.873\end{tabular}    & 
\begin{tabular}[c]{@{}c@{}}0.863\\ 0.894\\ 0.878\end{tabular}    
\end{tabular}
\end{table}

\subsubsection{Noisy Denial Constraints}\label{sec:noisy_dcs}
We now turn our attention to noisy constraints. We use the following definition of noisy constraints:
\begin{definition}
The denial constraint $dc$ is $\alpha$-noisy on the dataset $D$ if it satisfies $\alpha$ percent of all tuple pairs in $D$.
\end{definition}

\begin{figure*}
  \includegraphics[width=0.65\textwidth]{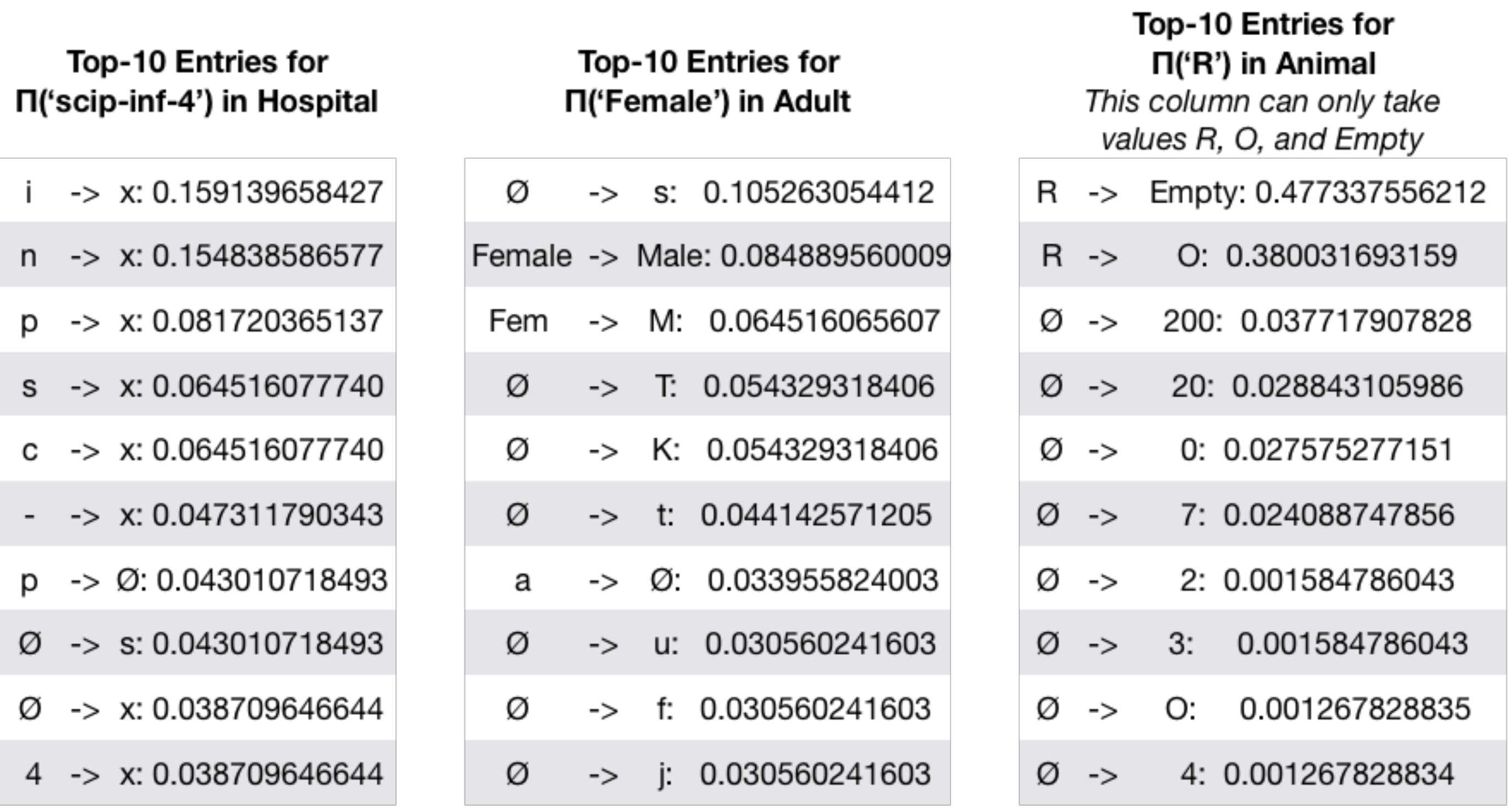}
  \caption{Examples of learned augmentation policies for clean entries in Hospital and Adult.}
  \label{fig:ex_policies}
\end{figure*}

We want to see the effect of noisy denial constraints on the performance of AUG. We use the following strategy to identify noisy denial constraints for each dataset: We use the denial constraint discovery method of Chu et al.~\cite{chu2013discovering} and group the discovered constraints in four ranges with respect to the noise level $\alpha$. Constraints with $\alpha \in (0.55, 0.65]$, constraints with $\alpha \in (0.65, 0.75]$, constraints with $\alpha \in (0.75, 0.85]$, and constraints with $\alpha \in (0.85, 0.95]$. For each range, we obtain 21 constraint-set samples, such that each sampled constraint set has the same cardinality as the original clean constraints associated with each of the Hospital, Adult, and Soccer datasets. We report the median performance of AUG in Table \ref{tab:noisydc}. As shown, the impact of noisy denial constraints on AUG's performance is not significant. The reason is that during training AUG can identify that the representation associated with denial constraints corresponds to a noisy feature and thus reduce its weight in the final classifier.

\begin{table}[]
\center
\scriptsize
\caption{Median performance of AUG over 21 runs with noisy constraints that correspond to different noise levels $\alpha$.}
\label{tab:noisydc}
\begin{tabular}{cc|cccc}
\textbf{Dataset}  & \textbf{M}                                        & \textbf{$\alpha\in(0.55,0.65]$}                                                  & \textbf{$(0.65,0.75]$}                                                  & \textbf{$(0.75,0.85]$}                                                  & \textbf{$(0.85,0.95]$}                                                    \\ \hline
\textbf{Hospital} & \begin{tabular}[c]{@{}c@{}}P\\ R\\ $F_1$\end{tabular} & \begin{tabular}[c]{@{}c@{}}0.859\\ 0.822\\ 0.840\end{tabular} & \begin{tabular}[c]{@{}c@{}}0.876\\ 0.869\\ 0.873\end{tabular} & \begin{tabular}[c]{@{}c@{}}0.912\\ 0.899\\ 0.906\end{tabular} & \begin{tabular}[c]{@{}c@{}}0.925\\ 0.914\\ 0.920\end{tabular} \\ \hline
\textbf{Adult}    & \begin{tabular}[c]{@{}c@{}}P\\ R\\ $F_1$\end{tabular} & \begin{tabular}[c]{@{}c@{}}0.911\\ 0.875\\ 0.893\end{tabular} & \begin{tabular}[c]{@{}c@{}}0.949\\ 0.930\\ 0.939\end{tabular} & \begin{tabular}[c]{@{}c@{}}0.961\\ 0.952\\ 0.956\end{tabular} & \begin{tabular}[c]{@{}c@{}}0.984\\ 0.961\\ 0.972\end{tabular} \\ \hline
\textbf{Soccer}   & \begin{tabular}[c]{@{}c@{}}P\\ R\\ $F_1$\end{tabular} & \begin{tabular}[c]{@{}c@{}}0.821\\ 0.864\\ 0.842\end{tabular} & \begin{tabular}[c]{@{}c@{}}0.849\\ 0.862\\ 0.855\end{tabular} & \begin{tabular}[c]{@{}c@{}}0.867\\ 0.880\\ 0.873\end{tabular} & \begin{tabular}[c]{@{}c@{}}0.863\\ 0.891\\ 0.877\end{tabular}
\end{tabular}
\end{table}

\subsection{Learned Augmentation Policies}
\label{sec:ex_policies}
We provide examples of learned policies for clean entries in Hospital, Adult, and Animal. For Hospital and Adult, we know how errors were introduced, and hence, can evaluate the performance of our methods for learning augmentation policies. Errors in Hospital correspond to typos introduced artificially by swapping a character in the clean cell values with the character `x'. On the other hand, errors in the gender attribute of Adult are introduced either by swapping the two gender values `Female' and `Male' or by introducing typos via injection of characters. For Animal, we do not know how errors are introduced. However, we focus on an attribute that can only take values in $\{$R, O, Empty$\}$ to evaluate the performance of our methods.

 Figure~\ref{fig:ex_policies} depicts the top-10 entries in the conditional distribution corresponding to entry `scip-inf-4' for Hospital and entry `Female' for Adult. As shown, for Hospital, almost all transformations learned by our method correspond to either swapping a character a character with the character `x' or injecting `x' in the original string. The performance of our approach is similar for Adult. We observe that a mix of value swaps, e.g., `Female' $\longmapsto$ `Male', and character injection transformations are learned. Finally, for Animal, we see that most of the mass of the conditional distribution (almost 86\%) is concentrated in the value swap transformations `R' $\longmapsto$ `Empty' and `R' $\longmapsto$ `O' while all other transformations have negligible probabilities. These results demonstrate that our methods can effectively learn how errors are introduced and distributed in noisy relational datasets.

\end{document}